\documentclass[11pt]{article}
\newcommand{\Comment}[1]{{}}

\usepackage{amsfonts,amsthm,amssymb,slashed}
\usepackage{amsmath}
\usepackage[textwidth = 430 pt, textheight = 630 pt]{geometry}
\usepackage{color} 
\usepackage{hyperref}
\usepackage{comment}
\usepackage{subfigure}
\usepackage{placeins}
\usepackage{cite}
\usepackage[compat=1.1.0]{tikz-feynman}
\usepackage{cancel}
\usepackage{geometry}
\usepackage{booktabs}
\usepackage{longtable}
\usepackage{array}
\usepackage{MnSymbol}
\usepackage{caption}
\usepackage{booktabs}

\Comment{\usepackage{color}
\definecolor{MyDarkBlue}{rgb}{0.15,0.15,0.45}
\usepackage[linktocpage=true]{hyperref}
\hypersetup{
colorlinks=true,
citecolor=MyDarkBlue,
linkcolor=MyDarkBlue,
urlcolor=MyDarkBlue,
pdfauthor={Authors },
pdftitle={Title},
pdfsubject={hep-th}
}

\usepackage[numbers,sort&compress]{natbib}
\usepackage{hypernat}}
\usepackage{graphicx}
\usepackage{cite}

\newcommand\ignore[1]{}
\def\one{{\,\hbox{1\kern-.8mm l}}}

\def\ra{\rangle}\def\la{\langle}

\newcommand{\pb}[1]{\bar{p}_{#1}}

\newcommand{\Cset}{{\,\,{{{^{_{\pmb{\mid}}}}\kern-.45em{\mathrm C}}}}}

\newcommand{\be}{\begin{equation}}
\newcommand{\bea}{\begin{eqnarray}}

\newcommand{\ee}{\end{equation}}
\newcommand{\eea}{\end{eqnarray}}

\definecolor{intensered}{RGB}{113, 0, 20} 

\begin{document}

\renewcommand{\thefootnote}{\fnsymbol{footnote}}

\makeatletter
\@addtoreset{equation}{section}
\makeatother
\renewcommand{\theequation}{\thesection.\arabic{equation}}

\rightline{}
\rightline{}




\begin{center}
{\LARGE \bf{\sc Cross-Correlations of Metric and Monopole Perturbations from Holographic Cosmology}}
\end{center} 
 \vspace{1truecm}
\thispagestyle{empty} \centerline{

{\large \bf {\sc Matheus Cravo${}^{a}$}}\footnote{E-mail address: \Comment{\href{mailto:matheus.cravo@unesp.br}}{\tt matheus.cravo@unesp.br}}
{\bf{\sc ,}} 
{\large \bf {\sc Juliana Z. Finotti${}^{a}$}}\footnote{E-mail address: \Comment{\href{mailto:jz.finotti@unesp.br}}{\tt jz.finotti@unesp.br}}
{\bf{\sc, and}}
{\large \bf {\sc Horatiu Nastase${}^{a}$}}\footnote{E-mail address: \Comment{\href{mailto:horatiu.nastase@unesp.br}}
{\tt horatiu.nastase@unesp.br}}
                                                        }

\vspace{.5cm}


\centerline{{\it ${}^a$Instituto de F\'{i}sica Te\'{o}rica, UNESP-Universidade Estadual Paulista}} 
\centerline{{\it R. Dr. Bento T. Ferraz 271, Bl. II, Sao Paulo 01140-070, SP, Brazil}}
\vspace{.3cm}

\vspace{1truecm}

\thispagestyle{empty}

\centerline{\sc Abstract}

In this paper, we present the 1-loop calculation of the three-point function $\langle TJJ \rangle$ 
of the stress-energy tensor and two insertions of $SO(3)$ global currents, using a 3d toy model 
for holographic cosmology. By applying the holographic dictionary that relates these QFT 
$n$-point functions to cosmological correlators, together with the $Sl(2,\mathbb{Z})$ duality 
that maps the electric Noether current to a magnetic vortex current dual to cosmological 
magnetic monopoles, we relate the $\langle TJJ \rangle$ correlator to the cross-correlations 
between metric perturbations and the bulk magnetic monopole field, specifically mapping to 
the non-Gaussianities $\langle \xi \tilde{A} \tilde{A} \rangle$ and $\langle \gamma \tilde{A} \tilde{A} 
\rangle$. We calculate the semi-local contact terms necessary to achieve the exact factorization of the 
cosmological correlators in the squeezed limit, $p_1 \to 0$. Finally, we evaluate the effective non-linear 
parameters, showing that the scalar-monopole cross-correlation vanishes at leading order, 
$f_{NL}^{\xi \tilde{A} \tilde{A}} = 0$, while the tensor-monopole cross-correlation yields a non-zero 
$f_{NL}^{\gamma\tilde A\tilde A}$. These results respect the expected amplitude hierarchy of 
the non-Gaussian correlators, while pointing at new directions in which holographic cosmology 
can be tested experimentally.
\vspace{.4truecm}

\begin{center}
\begin{minipage}[c]{380pt}
{\noindent 

}
\end{minipage}
\end{center}

\vspace{.5cm}

\setcounter{page}{0}
\setcounter{tocdepth}{2}

\newpage

\tableofcontents
\renewcommand{\thefootnote}{\arabic{footnote}}
\setcounter{footnote}{0}

\linespread{1.1}
\parskip 4pt
\newpage

\section{Introduction}

It has been a while since we have been trying to understand our universe
holographically. Following the successful conjecture of the
AdS/CFT correspondence in 1997, there have been a number of attempts
to find a similar relation for de Sitter space \cite{dscft,Maldacena,wittends,Susskind, holouniverse}, 
which is believed to be a good approximation for our universe, be it during the inflationary period or in the 
asymptotic future\footnote{Notice that recent DESI results put this fact in check \cite{DESI1,DESI2}.} 
\cite{musings,tasi}. This has not been an easy task for several reasons, including the difficulty of deriving 
de Sitter space top-down from string theory and the challenge of locating holographic degrees of freedom 
on a spacelike boundary that is only fully accessible to ``meta-observers"\footnote{For a review on dS 
holography, see \cite{musings}}. Alongside the more ambitious proposals, many works were and are 
being developed to explore specific features
of such a correspondence, e.g., attempting to build models of different
dimensionalities \cite{CSdS,microds3,vasiliev,SYK}, exploring the possible symmetries a dual theory
might have \cite{goheer,bootstrap,constraints}, or constraining and computing mathematical objects that
might be connected to observables in cosmology\cite{holouniverse,holonongauss}. 

One 
attempt states that de Sitter space must be dual to a conformal boundary theory, which is later 
formalized by the proposal that the Hartle-Hawking wavefunction of the universe is equal to the partition 
function of a CFT \cite{dscft,Maldacena}. A different attempt is the phenomenological approach of 
holographic cosmology by McFadden and Skenderis \cite{holouniverse,holocosmo}.
There, they establish a dictionary between correlation functions
of perturbative fields in inflationary cosmology
and $n$-point functions of boundary operators of a three-dimensional QFT. From this framework, it
was possible to compute the power spectra (2-point
function) of scalar $\left\langle \zeta\zeta\right\rangle $ and tensor
$\left\langle \gamma\gamma\right\rangle $ perturbations of the metric
field dual to the correlation function of the stress-energy tensor
$T_{\mu\nu}$ of the boundary theory. The results, when compared to
cosmological observations, provided a fit as good as the $\Lambda$CDM predictions 
\cite{wmap,planckdata}. More recently, it has been shown that this approach yields the same dictionary as 
the wavefunction one \cite{IRrenorm}.

Although not observed yet, other correlators have been computed, such as the bispectra (3-point 
functions) of metric perturbations\cite{holonongauss,holopredic,cosm3pt}, and two
and three-point functions of the magnetic monopole field \cite{nastase_pre-inflationary,matheuspaper}. 
The last two 
were part of finding a solution for the monopole problem within the holographic framework, which was 
done in \cite{nastase_pre-inflationary} alongside the resolution of other pre-inflationary
problems. 

More specifically, one computed on the boundary theory the 2- and 3-point
function of the global current that is dual to the magnetic monopole
field. The former gave us the anomalous dimension of this current, which
would explain the dilution of the monopole field throughout cosmic
evolution. The latter provides the non-Gaussian
distribution of the magnetic field. In this paper, we compute the
1-loop order of the $\left\langle TJJ\right\rangle $ correlator,
that corresponds to a cross-correlation between the metric perturbations
and the magnetic monopole in the bulk, namely $\left\langle \zeta \tilde{A} \tilde{A}\right\rangle $
and $\left\langle \gamma \tilde{A} \tilde{A}\right\rangle $. Yet still much weaker
than the 2-point function of the metric perturbations, these correlators
are expected to have greater amplitudes than $\left\langle \tilde{A}\tilde{A}\tilde{A}\right\rangle $(dual to $
\left\langle \tilde{J}\tilde{J}\tilde{J}\right\rangle $), making them the natural next observables to compute in 
this framework. Although none of these
are expected to be observed in the near future,
these computations are still interesting theoretically, considering
both the quantum field theory and bulk analyses. \footnote{Along this line of thought, one might ask why 
not compute $\left\langle TTJ\right\rangle $
first, since it is expected to have a greater amplitude than $\left\langle TJJ\right\rangle $.
We did, and at 1-loop order, it is trivially zero by the symmetry
of our model.}\textsuperscript{,}\footnote{The $\left\langle TTJ\right\rangle$ correlator has been computed in other contexts, such as in \cite{TJJfermion}, for Dirac fermions in 4D gravity.}

This paper is organized as follows. In \autoref{sec:review} we review the approach of holographic 
cosmology, its relation to the wavefunction approach, the solution to the monopole problem, and some 
remarks on non-Gaussianities. In \autoref{sec:calculation}, we compute the 3-point function for the stress-
energy tensor and the current operator and apply the holographic map to obtain the cosmological 
correlators, with special attention to the squeezed limit for the momenta. We conclude in 
\autoref{sec:discussion}, discussing what we found relevant in the process.

\section{Holographic cosmology, monopole problem and  non-\newline gaussianities}\label{sec:review}

\subsection{The phenomenological approach}
\label{section_pheno_holo_cosmo}

We now review the approach for holographic cosmology proposed in
\cite{holouniverse}. It assumes a phenomenological point of view of the subject,
rather than a top-down construction, that is, it is not a holographic
map derived through strings and brane dynamics like AdS/CFT. Instead,
one considers a more general model for the boundary QFT and set constraints
to it after comparison with observations. 

\begin{figure}[h]
    \centering
    \scalebox{0.6}{
    \begin{tikzpicture}[
        line width=1.0pt, 
        >={Stealth[length=3mm, width=2mm]},
        textnode/.style={align=center, font=\normalsize},
        arrowlabel/.style={align=center, font=\small}
    ]

        \filldraw[fill=white, draw=intensered, line width=1.8pt, join=round] (-1.8, 5.5) -- (-0.8, 6.5) -- (2.8, 6.5) -- (1.8, 5.5) -- cycle;
        \filldraw[fill=white, draw=intensered, line width=1.8pt, join=round] (1.8, 2.5) -- (2.8, 3.5) -- (2.8, 6.5) -- (1.8, 5.5) -- cycle;
        \filldraw[fill=white, draw=intensered, line width=1.8pt, join=round] (-1.8, 2.5) rectangle (1.8, 5.5);
        \node[textnode, intensered] at (0, 4) {inflationary\\[0.5ex]cosmology};

        \filldraw[fill=white, draw=black, line width=1.8pt, join=round] (6.2, 5.5) -- (7.2, 6.5) -- (10.8, 6.5) -- (9.8, 5.5) -- cycle;
        \filldraw[fill=white, draw=black, line width=1.8pt, join=round] (9.8, 2.5) -- (10.8, 3.5) -- (10.8, 6.5) -- (9.8, 5.5) -- cycle;
        \filldraw[fill=white, draw=black, line width=1.8pt, join=round] (6.2, 2.5) rectangle (9.8, 5.5);
        \node[textnode, black] at (8, 4) {domain\\[0.5ex]wall};

        \filldraw[fill=white, draw=intensered, line width=1.8pt, join=round] (-1.8, -3.5) rectangle (1.8, -0.5);
        \node[textnode, intensered] at (0, -2) {dual\\[0.5ex]pseudo-QFT};

        \filldraw[fill=white, draw=black, line width=1.8pt, join=round] (6.2, -3.5) rectangle (9.8, -0.5);
        \node[textnode, black] at (8, -2) {QFT};

        
        \draw[<->, black] (3.1, 4.3) -- (5.9, 4.3);
        \node[arrowlabel, black] at (4.5, 4.9) {DW/cosmology\\[0.2ex]correspondence};

        \draw[<->, black] (2.2, -2) -- (5.8, -2);

        \draw[<->, intensered, dashed, decorate, decoration={snake, amplitude=0.7mm, segment length=5mm, post length=3mm, pre length=3mm}] 
            (0, 2.1) -- (0, -0.1);

        \draw[<->, black, decorate, decoration={snake, amplitude=0.7mm, segment length=5mm, post length=3mm, pre length=3mm}] 
            (8, 2.1) -- (8, -0.1);
        \node[arrowlabel, black, right, xshift=2mm] at (8, 1.0) {gauge/gravity\\[0.2ex]duality};

    \end{tikzpicture}
    }
    \caption{Schematic figure illustrating the route taken by the framework \cite{holouniverse} to map an 
    accelerating universe to a quantum field theory description. The first step starts from a cosmological 
    spacetime and ends on a domain wall spacetime by a correspondence between the two. The second 
    step is the holographic description of the domain wall, leading to a 3-dimensional QFT. The last step is 
    to apply the inverse analytic continuation used in the first step, but now in QFT variables, leading to 
    QFT correlators dual to cosmological observables. The steps in black define the putative holographic 
    map, displayed in red, between our universe and a dual quantum field theory.}
    \label{figure}
    \end{figure}
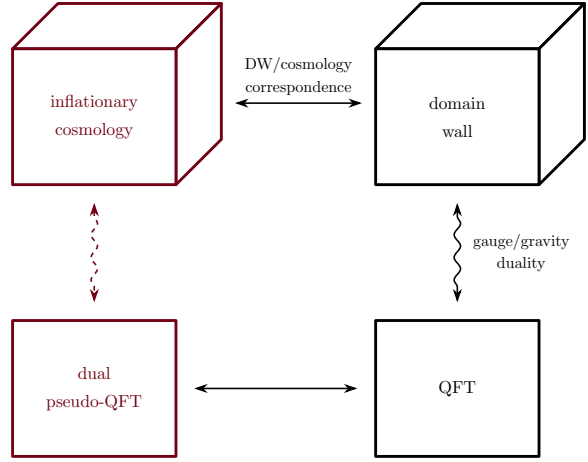

Although there is no direct duality between our universe and a QFT,
this approach revolves around an one-to-one correspondence between
($d$-dimensional) cosmologies and domain-wall spacetimes \cite{dwcosmo1,dwcosmo2},
which, in turn, have a dual description in terms of a $(d-1)$-dimensional
QFT. The domain wall/cosmology map is also applicable to QFT variables,
so one can then obtain the QFT dual to the original cosmological spacetime.
These steps are illustrated in \autoref{figure}.
    
The first step in \autoref{figure} represents the cosmology/domain wall
correspondence, which is roughly an analytic continuation\footnote{More precisely, 
a {\em correspondence} between quantities in one theory and in another; a naive analytical continuation
would lead to complex (not just imaginary) quantities.} of their
solutions. A flat D-dimensional domain wall is described by the metric
\begin{equation}
ds_{DW}^{2} = dz^{2}+e^{2A(r)}dx^{2},
\end{equation}
 where $z$ is the radial coordinate (perpendicular to the slice)
and $A(r)$ determines the geometry of the $(D-1)$-dimensional slice.
If we consider the system 
\begin{equation}
S=\frac{1}{2\kappa^{2}}\int d^{4}x\sqrt{\left|g\right|}\left[-R+\frac{1}{2}\left|\partial\Phi\right|^{2}
+2\kappa^{2}V(\Phi)\right]
\end{equation}
with a monotonic scalar field $\Phi(z)$, the correspondence states
that the same model with potential $-V$ has the  \emph{cosmological}
solution
\begin{equation}
ds^{2}=-dt^{2}+2e^{2A(t)}dx^{2},
\end{equation}
and a scalar field that depends only on time, $\Phi=\phi(t)$. For
$A(t)=\ln a(t)$, this is the Friedmann-Lemâitre-Robertson-Walker
(FLRW) metric. 

The connection between the two solutions is the analytical continuation
$t\rightarrow it$ after defining $t=iz$. This can also be extended to linear perturbations of the 
background solutions, following the
map 
\begin{equation}
    \bar{\kappa}^{2} = -\kappa^{2}, \qquad \bar{p} = -i p, \label{analytical_continuation_dw_cosmo}
\end{equation}
where $\kappa$ is Einstein's constant. This way, cosmological perturbations parameterized by 
$\left(\kappa,p\right)$ are mapped to perturbed domain-wall solutions parameterized by $
\left(\bar{\kappa},\bar{p}\right)$. Later, we will go back to cosmology side, as shown in the last step of \autoref{figure}. In QFT variables, this means going back to momentum $p=i\bar{p}$ and taking the 
rank of the gauge group of the pseudo QFT dual to cosmology to be $N^2 =- \bar{N}^2$.

The next step is to describe the domain-wall solutions holographically. Luckily, these spacetimes have 
such a description \cite{dwholography}, so we can map these solutions to a quantum field theory in one 
dimension less. Under gauge/gravity dualities, bulk fields $\Phi$ are dual to gauge-invariant operators in 
the boundary theory and, in particular, the metric in the gravity
side couples to the energy-momentum tensor of the QFT. Ultimately,
one can write the power spectra associated to metric fluctuations
in terms of the 2-point function of the energy-momentum tensor of
the dual field theory,
%
%
\begin{equation}
    \Delta^{2}_{S}(p)=-\frac{q^{3}}{16\pi^{2}}\frac{1}{\text{Im}\bar{B}(\bar{p})}\Bigg|_{\text{analyt. cont.}},
    \qquad \quad \Delta^{2}_{T}(p)=-\frac{2q^{3}}{\pi^{2}}\frac{1}{\text{Im}\bar{A}(\bar{p})}\Bigg|
    _{\text{analyt. cont.}},\label{holographic_formulas_scalar_tensor_power_spectrum}
\end{equation}
where $\bar A (\bar p)$ and $\bar B(\bar p)$ are the coefficients of the decomposition of the energy-
momentum tensor correlator into its Lorentz structures 
\begin{equation}
\llangle T_{\mu\nu}\left(\bar{p}\right)T_{\rho\sigma}\left(-\bar{p}\right)\rrangle = \bar A(\bar{p})
\Pi_{\mu\nu\rho\sigma} + \bar B(\bar{p})\pi_{\mu\nu}\pi_{\rho\sigma}, 
\label{tensor_decomposition_2pt_function_TT}
\end{equation}
with 
\begin{equation}
\Pi_{\mu\nu\rho\sigma}=\frac{1}{2}\left(\pi_{\mu\rho}\pi_{\nu\sigma}+\pi_{\mu\sigma}\pi_{\rho\nu}-
\pi_{\mu\nu}\pi_{\rho\sigma}\right), \quad \text{and} \quad \pi_{\mu\nu} = g_{\mu\nu} - \frac{\bar p_\mu \bar 
p_\nu}{\bar p^2},
\end{equation}
for $d=3$, and the analytical continuation to cosmology means 
\begin{align}
    \frac{1}{\text{Im}\bar{A}(\bar{p})}\Bigg|_{\text{analyt. cont.}}&=\frac{1}{\text{Im}\bar{A}(\bar{p})}\Bigg|
    _{\bar{p}=-ip,\;\;\bar{N}=-iN}\equiv\frac{1}{\text{Im}A(-ip)},
    \\\frac{1}{\text{Im}\bar{B}(\bar{p})}\Bigg|_{\text{analyt. cont.}}&=\frac{1}{\text{Im}\bar{B}(\bar{p})}\Bigg|
    _{\bar{p}=-ip,\;\;\bar{N}=-iN}\equiv\frac{1}{\text{Im}B(-ip)},
\end{align}
at the end of the calculation. Note that in \eqref{tensor_decomposition_2pt_function_TT} we are using 
the double bracket notation to denote the delta-stripped correlator,
\begin{equation}
  \langle \mathcal{O}_1(\bar p_1) \mathcal{O}_2(\bar p_2) \rangle = (2\pi)^3 \delta^{(3)} (\bar p_1 
  + \bar p_2)\llangle \mathcal{O}_1(\bar p) \mathcal{O}_2(-\bar p)\rrangle\;. 
  \label{notation_correlators_llangle_rrangle}
\end{equation}

This dictionary is established from the point of view of semiclassical gravity in the bulk, because it is 
built considering the asymptotic behaviour of metric perturbations. However, the map is valid at all scales 
in both theories and, in particular, it will be the most useful when gravity is strongly coupled in the bulk, 
that is, at early times in the cosmological evolution. Then, we will be able to do calculations in a weakly 
coupled QFT, and those will carry information about a non-geometric phase in the bulk.

As this is a phenomenological approach and not a top-down construction,
we must write a general action for the dual (to the DW) field theory.
Although we don't know its exact composition, we know that it must
be super-renormalizable, admit a large $\bar{N}$ limit of its $SU(\bar{N})$
gauge group and be equipped with a generalized conformal structure.\footnote{This is a feature of 
theories obtained upon dimensional reduction of a CFT. In fact, it behaves as a CFT if one considers the 
Yang-Mills constant $g_{\text{YM}}$ to transform as a scalar field. See \cite{gcs1,gcs2}.}

The 3-dimensional Euclidean model proposed to be the domain wall gravity dual is
\begin{align}\label{class_models_holo}
    S &=\frac{2}{g_{YM}^{2}}\int d^{3}x\text{Tr} \left[\frac{1}{4}F_{\mu\nu}^{I}F_{\mu\nu}^{I}+\frac{1}
    {2}\left(D\Phi^{J}\right)^{2}+\bar{\Psi}^{L}\cancel{D}\Psi^{L} \right. \nonumber \\
    & \qquad \left. +\lambda_{M_{1}M_{2}M_{3}M_{4}}\Phi^{M_{1}}\Phi^{M_{2}}\Phi^{M_{3}}\Phi^{M_{4}} +
    \mu_{ML_{1}L_{2}}^{\alpha\beta}\Phi^{M}\Psi_{\alpha}^{L_{1}}\Psi_{\beta}^{L_{2}}\right],
\end{align}
where one considers $\mathcal{N_{A}}$ gauge fields, $\mathcal{N}_{\Phi}$ scalars and $\mathcal{N}
_{\Psi}$ fermions, such that 
\begin{equation}
I=1,...,\mathcal{N}_{A}; \qquad M=1,...,\mathcal{N}_{\Phi}; \qquad L=1,...,\mathcal{N}_{\Psi};
\end{equation}
are the flavor indices and all fields transform in the adjoint representation $\Phi=\Phi^aT_a$, where $T_a$ 
are the generators of $SU(N)$, normalized by ${\rm Tr} (T^a T^b)=\frac{1}{2}\delta^{ab}$.

The interactions of the model are a Yukawa and a quartic-scalar terms, with coupling constants $\mu$ 
and $\lambda$, respectively. Once one writes the Yang-Mills coupling $g_{\text{YM}}$ as an overall factor, 
the other couplings become dimensionless, and the fields have dimensions as if they were in a 4-
dimensional theory. In fact, this is a property of the generalised conformal structure, which is not a 
symmetry, but still it constrains the form of the correlators of the theory such that, in the perturbative 
regime, they can be expanded as a function of an effective dimensionless coupling 
\begin{equation}
  g^2_{\text{eff}}=\frac{g_{\text{YM}}^2 \bar N}{\bar p}.
\end{equation}

We can write the coefficients $A(\bar{p})$ and $B(\bar{p})$ of the 2-point function 
\eqref{tensor_decomposition_2pt_function_TT} as 
\begin{equation}
  A(\bar{p})=\bar{p}^3 \bar{N}^2 f_T(g^2_{\text{eff}}), \hspace{2cm} B(\bar{p})=\frac{1}{4}\bar{p}^3 \bar{N}
  ^2 f(g^2_\text{eff}),
\end{equation}
where $\bar{p}^3$ comes from the overall dimensional factor $p^{2\Delta -d}$, where $\Delta = 3$ is the 
classical dimension of the energy-momentum tensor in $d=3$, and $\bar{N}^2$ is the contribution for the 
contraction of all $SU(\bar{N})$ indices for the bubble planar diagrams. Generalized conformal symmetry 
fixes the form of the functions $f_T(g^2_{\text{eff}})$ and $f(g^2_\text{eff})$ to be, perturbatively,
\begin{align}
  f(g^2_\text{eff}) &=f_0 \left[ 1- f_1 g^2_\text{eff} \ln (g^2_\text{eff}) + f_2 g^2_\text{eff} +\mathcal{O}
  (g^4_{eff}) \right], \label{f_exp}\\
  f_T(g^2_\text{eff}) &=f_{T0} \left[ 1- f_{T1} g^2_\text{eff} \ln (g^2_\text{eff}) + f_{T2} g^2_\text{eff} +
  \mathcal{O}(g^4_{eff}) \right], \label{fT_exp}
\end{align}
where $\{f_0,f_{T0}\}$ and $\{f_1,f_{T1},f_2,f_{T2}\}$ are the one-loop and two-loop contributions, 
respectively.

Finally, having done computations in the domain-wall QFT, the last step in \autoref{figure} consists in 
performing the analytic continuation ``back" to cosmology, this time in QFT variables. After applying it to 
\eqref{holographic_formulas_scalar_tensor_power_spectrum}, we find the cosmological power spectra to 
be
\begin{equation*}
  \Delta_{S(T)}^{2}(q)=\frac{\Delta_{0(T0)}^{2}}{\left[1+\left(g_{(T)}\frac{p_{*}}{p}\right)\ln\left|\frac{p}{\beta_{T} g_{(T)}p_{*}}\right|+\mathcal{O}\left(g_{(T)}\frac{p_{*}}{p}\right)^{2}\right]},
\end{equation*}

where $\Delta_{0(T)}$, $g_{(T)}$ and $\beta_{(T)}$ are parameters defined in terms of the coefficients of 
the functions $f_{(T)}(g_{\text{eff}}^2)$ as
\begin{align}
  g p_* &= f_{1}g_{\text{YM}}^2 N ,\\
  \ln \frac{1}{\beta} &= \frac{f_{2}}{f_{1}}+ \ln\left|f_{1} \right|,\\
  \text{and }\Delta_{0}^{2} &=\frac{1}{4\pi^{2}N^{2}f_{0}}.
\end{align}

Notably, the functional form of the power spectrum derived from the dual QFT approach is different from 
that typically derived by inflationary models. And yet, after comparisons of both predictions with CMBR 
data, it was found that the two have a good fit\cite{wmap,planckdata}. In fact, the difference between them 
is statistically insignificant! Moreover, analysis of these data has been used to fix some features of the 
phenomenological QFT. For instance, a gauge theory coupled to only
fermions is ruled out, and a Yang-Mills theory with a large number of nearly conformal scalars and quartic 
interactions is preferable.
%
%
%
%
%
\subsection{Monopole problem}

\label{section_monopole_problem}

If this framework aims to be a good description of cosmology at early times, it must solve the pre-
inflationary problems. Indeed, it does, as it is discussed in detail in \cite{nastase_pre-inflationary}. Among 
these problems, the monopole problem and how it is understood on the QFT side are of special interest to 
us. 

The monopole problem evolves around the fact that broken symmetries in the GUT phase transition 
predict a very high density of magnetic monopoles created at that moment. These are non-trivial 
topological solutions of the field equations of a $SO(3)$ gauge theory that is spontaneously broken into a 
$U(1)$ symmetry (for a review, see \cite{preskill}). It was shown by 't Hooft and Polyakov that such GUT 
theories necessarily contain magnetic monopoles. However, none has been observed yet, and even if it 
had been, the predicted density is so high (the estimated monopole mass would be of the order of 
$10^{16}GeV$, and the Kibble mechanism would dictate to have about one monopole
per horizon at the phase transition time) that it would lead the universe to collapse gravitationally. 

The usual solution for this problem is to propose a mechanism by which this density is diluted throughout 
the evolution of the universe. Inflationary models achieve that through the period of exponential inflation 
by requiring a minimal number of e-folds ($N_{e}>\ln10^{10}\simeq23$). 

In holographic cosmology, this is solved from the point of view of the dual field theory. Under gauge/
gravity duality, a (topological) monopole field $\tilde{A^{\mu}}$ in the bulk couples to a (vortex) magnetic 
current $\tilde{J_{\mu}}$ in the boundary. Then, if the monopole density is small at late times, which 
means that the bulk operator dies off there, conversely, the magnetic current $\tilde{J_{\mu}}$ must be a 
relevant operator in the RG flow of the dual QFT. This translates into $\tilde{J_{\mu}}$ having an 
anomalous dimension $\tilde{\delta}<0$, information that can be extracted from the momentum 
dependence of the two-point function $\llangle \tilde{J_{\mu}}(\bar q)\tilde{J_{\nu}}(-\bar q) \rrangle$. 
Indeed, it has, at least for some bosonic toy models in the phenomenological class of models 
\eqref{class_models_holo}. Let us briefly review this solution.

First, one needs to specify a model to do this calculation over. We cannot consider the general 
phenomenological model because we are interested in a model with a particular symmetry that admits a 
specific solution. That is, a model with $SO(3)$ global symmetry with vortex solutions. The reason for that 
is that we need a boundary theory with the global symmetry and pattern of symmetry breaking that are 
analogous to the creation of monopoles in the bulk theory. It follows that the vortex solution is analogous 
to the hedgehog configuration in the 't Hooft monopole problem, in the sense that it minimizes the 
potential energy with a non-trivial topology and is $U(1)$ invariant.

The toy model proposed in \cite{nastase_pre-inflationary} is the simplest one with gauge fields and 
scalars, since models without fermions are preferred according to the fitting with CMBR data.  As already 
said, we want a theory that preserves the global symmetry, for that we have six ($i=1,2$) complex scalars 
$\phi_{i}^{a}$ in the triplet representation of $SO(3)$ ($a=1,2,3$), and the quartic potential 
$\left|\vec{\phi_{1}}\times\vec{\phi_{2}}\right|^{2}$. Lastly, all the fields transform in the adjoint 
representation of $SU(N)$, but we keep these indices implicit. The action in Euclidean space is 
\begin{equation}\label{toymodel}
S=\frac{2}{g_{YM}^{2}}\int d^{3}x \,\text{Tr}\left[\frac{1}{4}F_{\mu\nu}F^{\mu\nu}+\left|D_{\mu}\vec{\phi_{i}}
\right|^2+\lambda\left|\vec{\phi_{1}}\times\vec{\phi_{2}}\right|^{2}\right],
\end{equation}
and the global Noether current is 
\begin{equation}\label{noether}
J_{\mu}^{a}=\sum_{i=1,2}\vec{\phi_{i}^{*}}T^{a}D_{\mu}\vec{\phi_{i}} + \text{h.c.}\,
\end{equation}
where $T^{a}$ are the $SO(3)$ generators, and the field $\vec{\phi}$ is a $SO(3)$ triplet with components 
$\phi^{a}$. 

As we are interested in the dilution of the monopole density, we want to look at what happens to the 
magnetic (vortex) current $\tilde{J_{\mu}}$, dual to the monopole configuration $\tilde{A}_{\mu}$, in the 
UV regime. This information is encoded in its anomalous dimension, which can be extracted from the 2-
point function $\llangle \tilde{J}_{\mu}(\bar q)\tilde{J}_{\nu}(- \bar q)\rrangle$. However, when we consider 
a global symmetry, it is natural to work with the Noether current $J_{\mu}$ \eqref{noether} which 
couples to 
the electromagnetic field. Luckily, the 2-point function of the vortex current is related to that of the electric 
current by what is the three-dimensional version of the electric-magnetic duality. Following the 
analysis in \cite{wittenSL2,particlevortex,Herzog}, the current correlator is written as
\begin{equation}
\llangle J_{\mu}(\bar p)J_{\nu}(-\bar p)\rrangle =\left(\bar p^{2}\delta_{\mu\nu}-\bar p_{\mu}\bar p_{\nu}\right)\frac{t}{2\pi\sqrt{\bar p^{2}}}+\epsilon_{\mu\nu\rho}\bar p_{\rho}\frac{\omega}{2\pi}, \label{2pt_function_JJ_t_omega}
\end{equation}
where $t$ and $\omega$ are functions of the couplings of the theory. In our case, for instance, 
$\omega=0$. Then, the $Sl(2;\mathbb{Z})$ responsible for action on the electric and magnetic 
currents acts on $t$ and $\omega$ as
\begin{equation}
t\rightarrow\frac{t}{t^{2}+\omega^{2}},\ \ \ \ \ \ \!\ \text{and}\ \ \ \ \ \ \ \!\ \omega\rightarrow\frac{\omega}
{t^{2}+\omega^{2}}, \label{Sl2Z_duality}
\end{equation}
such that, for $\omega=0$, this amounts to the inversion $t\rightarrow1/t$. Then, the magnetic current 
correlator is
\begin{equation}
\llangle \tilde{J}_{\mu}(\bar p)\tilde{J}_{\nu}(-\bar p)\rrangle = \frac{\bar p}{2\pi t}\left(\delta_{\mu\nu}-
\frac{\bar p_{\mu}\bar p_{\nu}}{\bar{p}^{2}}\right).
\end{equation}

Note that this was developed for conformal field theories. When we generalise it to a theory with a general 
conformal structure, the parameter $t$ is now a function of the effective coupling constant $g_{\text{eff}}$. 
Therefore, when computing the 2-point function, we can expand it as $t\simeq t_{0}(1+ \delta \log \bar p)$ 
to extract the anomalous dimension, and we see that the inversion of $t$ turns it into $t\simeq t_{0}(1-
\delta \log \bar p)$. This means that the anomalous dimension of the vortex current is related to the 
electric current by $\tilde{\delta}(\tilde{J})=-\delta(J)$. Therefore, if we find the electric current to be an
irrelevant operator, then the vortex current is relevant, and the dual monopole field dies off at late times in 
cosmology, promoting the dilution of the monopole density throughout cosmological evolution.

Performing the computation of the two-point function of the Noether current $J^a_\mu$ at two loops, it 
was found that its anomalous dimension is
\begin{equation}
\delta=\frac{8}{\pi^2}\frac{g_{\text{YM}}^2 \bar N}{\bar p}.
\end{equation}

This anomalous dimension is positive and invariant under the analytical continuation $\bar p = -i p$, and 
$\bar N = -i N$ (or for $\bar p = ip$ and $\bar N = i N$, as discussed before), meaning that $J^a_\mu$ is 
an irrelevant operator both in the domain wall and cosmology variables. Consequently, the vortex operator 
$\tilde{J^a_\mu}$ is a relevant operator, and the monopole problem is solved. Moreover, this result was 
found to be independent of the explicit form of the potential, because all Feynman
diagrams containing the quartic vertex were zero in dimensional regularization. Then, despite having used 
a specific toy model to do this calculation, the solution of the monopole problem can be extended to a 
whole class of models, provided they contain only bosons and a potential that allows vortex solutions.
\subsection{Holographic cosmology from the wavefunction of the Universe}

An alternative approach to obtain the holographic formulas 
\eqref{holographic_formulas_scalar_tensor_power_spectrum} is to consider Maldacena's prescription 
\cite{Maldacena} relating the wavefunction of the Universe (originally for de Sitter space) and the partition 
function of a conformal field theory:
\begin{equation}
  \Psi_{\text{dS}}[h, \varphi] = Z_{\text{CFT}}[h,\varphi] \;,
\end{equation}
where the sources $h, \varphi$ are, in the context of cosmology, the late-time values of the bulk fields. 
This relation is understood as an analytical continuation of the AdS/CFT dictionary, or more precisely, 
one uses the fact that AdS and dS are related by $L_{\text{AdS}} \rightarrow iL_{\text{dS}}$ and 
$z \rightarrow 
-it$, where $L_{\text{AdS}}$ ($L_{\text{dS}}$) is the AdS (dS) radius. This prescription gives the same 
holographic formulas as the holographic cosmology using the domain wall cosmology correspondence, as 
discussed in detail in \cite{IRrenorm}, and here we extrapolate it to the non-geometric phase as well. The 
only caveat is that, as discussed in \cite{IRrenorm}, the analytical continuation 
\eqref{analytical_continuation_dw_cosmo} actually corresponds to $\Psi^* = Z$, and instead, we are 
going to use 
\begin{equation}
    \bar N = i N, \qquad \bar p = i p, \label{analytical_continuation_WF_QFT}
\end{equation}
which corresponds to the mapping $\Psi = Z$. Both prescriptions reproduce the same results, and we 
will adopt the latter to keep consistency with the more recent literature. In these conventions, the 
holographic formulas for 2- and 3-point functions are 
\begin{align}
 \llangle \varphi(p)\varphi(-p) \rrangle & = -\frac{1}{2} \frac{1}{\operatorname{Re}[(-i)^d \llangle \mathcal{O}
 (\bar p)\mathcal{O}(-\bar p) \rrangle]}\Bigg|_{\bar{p}=ip,\;\;\bar{N}=iN}, 
 \label{holographic_formula_2pt_function} \; \\
  \llangle \varphi(p_1)\varphi(p_2)\varphi(p_3) \rrangle & = \frac{1}{4} \frac{\operatorname{Re}[(-i)^d 
  \llangle \mathcal{O}(\bar p_1)\mathcal{O}(\bar p_2)\mathcal{O}(\bar p_3) \rrangle]}{\prod_{i=1}^3 
  \operatorname{Re}[(-i)^d \llangle \mathcal{O}(\bar p_i)\mathcal{O}(-\bar p_i) \rrangle]}\Bigg|_{\bar{p}=ip,
  \;\;\bar{N}= iN} \; .\label{holographic_formula_3pt_function}
\end{align}
These formulas were derived for scalar cosmological observables, so we can formally write the scalar 
QFT correlators in the denominator of the right-hand side. 
For operators and sources with Lorentz indices, on the 
other hand, the direct generalization is 
\begin{equation}
    \llangle\varphi_{\{i_{1}\}}(p_{1})\varphi_{\{i_{2}\}}(p_{2})\varphi_{\{i_{3}\}}(p_{3})\rrangle = 
    \frac{\mathrm{Re}\left[(-i)^{d}\llangle\mathcal{O}^{\{j_{1}\}}(\bar p_{1})\mathcal{O}^{\{j_{2}\}}(\bar p_{2})
    \mathcal{O}^{\{j_{3}\}}(\bar p_{3})\rrangle\right]}{4\prod^{3}_{k=1}\mathrm{Re}\left[(-i)^{d}
    \llangle\mathcal{O}^{\{i_{k}\}}(\bar p_{k})\mathcal{O}^{\{j_{k}\}}(-\bar p_{k})\rrangle\right]} \Bigg|_{\bar{p}
    =ip,\;\;\bar{N}=iN} ,\label{holographic_formula_3pt_function_tensor}
\end{equation}
where ${i}$ denotes a set of Lorentz indices, and repeated sets of indices mean full contraction 
(up to symmetry factors). Here, the QFT correlators in the denominator actually mean the 
\emph{inverse} of the tensor 2-point functions, that is, 
\begin{equation}
     \frac{1}{ G^{\{ij\}}} \equiv (G^{\{ij\}})^{-1}
\end{equation}
Finding this inverse may not be trivial, as we will see.
\subsubsection{Cosmological non-Gaussianities} \label{cosmonongauss}

So far, we have explored the holographic point of view of primordial cosmological perturbations up to 
linear order, that is, of Gaussian perturbations. These arise from the behaviour of free fields; they are 
responsible for structure formation in our universe and are imprinted in the cosmic microwave 
background.

Conversely, non-Gaussian perturbations represent the deviation from perfect Gaussian statistics. These 
lead to non-zero higher order correlators and evaluate the effect of interactions between primordial fields. 
Although we have not observed non-Gaussianities yet, we may still go further and use the holographic 
cosmology framework to predict higher-order cosmological correlators.

For a generic set of scalar cosmological observables $\varphi_i(x)$, the three-point function is defined as 
\begin{equation}
    \langle \varphi_1(p_1) \varphi_2(p_2) \varphi_3(p_3) \rangle = (2\pi)^d \delta^{(d)}(p_1 + p_2 + p_3) 
    \mathcal{B}_{\varphi_1 \varphi_2 \varphi_3} \, , 
\end{equation}
where the overall amplitude $\mathcal{B}_{\varphi_1 \varphi_2 \varphi_3}$ is called the 
\emph{bispectrum} of the corresponding non-Gaussianity. The functional dependence of the bispectrum 
on the configuration of the momentum triangle determines the ``shape'' of the non-Gaussianity. Among 
the various configurations, the squeezed limit (or soft limit) is defined as the kinematic regime where one 
of the momenta is much smaller than the other two. Taking $p_1 \to 0$, momentum conservation requires 
$p_2 \approx - p_3 \equiv p$. 

During inflation, modes with different momenta exit the Hubble horizon at different times. A soft mode 
$p_1 \to 0$ corresponds to a long-wavelength fluctuation that crosses the horizon and freezes out before 
the short-wavelength modes $p$. In this regime, the long-wavelength mode has already become classical 
and acts as a constant, classical background field. 

Suppose we have the cross correlation between an observable $\varphi_1$ and two insertions of another 
observable $\varphi_2$. In the squeezed limit with the momentum associated with $\varphi_1$ taken to 
zero, the three-point function factorizes as 
\begin{equation}
   \lim_{p_1 \rightarrow 0} \llangle \varphi_1(p_1) \varphi_2(p) \varphi_2(-p) \rrangle \sim \left(\lim_{p_1 
   \rightarrow 0} \mathcal{P}_{\varphi_1}(p_1) \right) \mathcal{D}(p) \mathcal{P}_{\varphi_2}(p) \, ,
   \label{factorization_squeezed_limit}
\end{equation}
where $\mathcal{D}(p)$ captures how the two-point function of the $\varphi_2$ field reacts to the 
background modulation given by the insertion of $\varphi_1$, and it is typically proportional to the 
logarithmic derivative of the hard field's power spectrum and ${\cal P}$ refers to the 
2-point function (power spectrum). Being specific, if $\varphi_1$ is the scalar 
perturbation $\xi$ and $\varphi_2$ is the tensor perturbation (for a given polarization $s_i$), then the 
three point function $\langle \xi \gamma \gamma \rangle $ is given by 
\begin{align}
    \lim_{p_1 \to 0} \llangle \xi(p_1) \gamma_{s_2}(p_2) \gamma_{s_3}(p_3) \rrangle &= P_\xi(p_1) 
    P_{\gamma}(p) \left[ - \frac{d \ln P_{\gamma}(p)}{d \ln p} \delta_{s_2 s_3} \right] \nonumber \\ 
    &  = - P_\xi(p_1) \frac{d}{d \ln p} P_{\gamma}(p) \delta_{s_2 s_3}\, .
\end{align}
In this specific example, the three-point function measures the correlation between the long mode 
associated with the scalar perturbation and the power spectrum of the short modes for the tensor 
perturbation. Because the soft mode merely alters the background on which the short modes evolve, the 
effect of the interaction can be derived completely from symmetry arguments. Specifically, a soft scalar 
mode $\xi(p_1 \to 0)$ induces a local spatial conformal transformation (a rescaling of the spatial 
coordinates). If instead we calculate $\langle \gamma \xi \xi\rangle$ in the squeezed limit, the 
interpretation is that the soft tensor mode $\gamma(p_1 \to 0)$ induces an anisotropic shear.

To quantitatively compare the theoretical predictions of different models with observational constraints, we 
introduce the dimensionless non-linear parameter, $f_{NL}$. This parameter normalizes the amplitude of 
the bispectrum relative to the square of the power spectrum. The original definition of $f_{NL}$ is based 
on the local shape of non-Gaussianity, which peaks precisely in the squeezed limit. Phenomenologically, 
local non-Gaussianity is defined by an expansion of the perturbation around a perfectly Gaussian field. 
For the scalar perturbation, for instance, we have an expansion around $\xi_G(\mathbf{x})$:
\begin{equation}
    \xi(\mathbf{x}) = \xi_G(\mathbf{x}) + \frac{3}{5} f_{NL}^{\text{local}} \left[ \xi_G^2(\mathbf{x}) - \langle 
    \xi_G^2 \rangle \right] \, .
\end{equation}

We can extend this definition to mixed correlators, which requires handling the dimension of operators 
and tensor contractions. For a cross-correlator like $\langle \xi \chi \chi \rangle$, an effective mixed 
non-linear parameter $f_{NL}^{\text{mixed}}$ can be defined in many different ways in terms of the 
crossed bispectrum and the dimensionless power spectrum, and it higly depends on the "shape" of the 
momentum triangle. Here we define the non-linear coupling $f_{NL}^{\text{mixed}}$ in the squeezed limit 
by the consistency relation:
\begin{equation}\label{def_shapefunction}
    B_{\xi \chi \chi}(p_1, p, p) \sim f_{NL}^{\text{mixed}} \, \mathcal{P}_{\xi}(p_1) \mathcal{P}_{\chi}(p) \times 
    (\text{kinematic factors}) \, ,
\end{equation}
which means that in this setup, 
\begin{equation}
    D(p) = f_{NL}^{\text{mixed}}(p)\, . \label{equality_D_f_NL}
\end{equation}

Now we turn our attention back to holographic cosmology. The scalar bispectrum is dual to the trace of 
the stress-energy tensor of the dual QFT. Using the phenomenological models for holographic cosmology 
we discussed in \autoref{section_pheno_holo_cosmo}, it was found that the cosmological bispectrum has 
only an equilateral form, with $f^\text{equil}_\text{NL}=5/36$. The shape function was found to be 
completely independent of the field content of the dual QFT, as they appear as a multiplicative factor in 
the correlator \cite{holonongauss}. Holographic 3-point functions involving also tensor perturbations of the 
metric were derived and computed in \cite{cosm3pt,holopredic}.

Besides the interest in the non-Gaussianities of the scalar and tensor perturbations, we can also ask the 
question of how the power spectrum of the monopole field $\tilde A$ is affected by the modulation of 
scalar and tensor perturbations in the squeezed limit. After the computation of the 2-point function of the 
current operator, the following step is 
to compute its 3-point function. Under the holographic map, these account 
for non-Gaussianities in the monopole distribution, and were calculated in \cite{matheuspaper}.

The next step is to compute the $\left<TJJ \right>$ correlator, which corresponds to a cross-correlation 
between primordial perturbations of the metric and the monopole field. As discussed in the introduction, 
there are two cosmological correlators we want to calculate holographically: $\la \xi \tilde A \tilde A \ra$ 
and $\la \gamma \tilde A \tilde A \ra$. The precise map between cosmological observables (sources) and 
gauge-invariant operators on the QFT side for this case is 
\begin{equation}
    \xi(p) \leftrightarrow T(p), \qquad \gamma_{\mu \nu}(p) \leftrightarrow\frac{1}{4}T^{\text{TT}}_{\mu \nu} 
    (p), \qquad \tilde A^a_\mu(p) \leftrightarrow \tilde J^a_\mu(p) \,,
\end{equation}
where the trace of the energy momentum is $T(p) = \delta^{\mu \nu} T_{\mu \nu}(p)$ and its 
transverse-traceless projection is given by $T^{TT}_{\mu \nu}({p})
=\Pi_{\mu \nu}{}^{\rho \sigma}T_{\rho \sigma}(p)$. 
The magnetic vortex current $\tilde J_\mu^a$ is the same as we introduced in the 
\autoref{section_monopole_problem}.

Using the formula \eqref{holographic_formula_3pt_function_tensor}, we start by evaluating the 
cosmological correlator
\begin{align}
& \llangle\xi(p_{1})\tilde{A}^a_{\rho}(p_{2})\tilde{A}^b_{\sigma}(p_{3})\rrangle  = \nonumber \\
& \frac{\mathrm{Re}\left[(-i)^{d}\llangle T(\bar{p}_{1})\tilde{J}^{c\,\lambda_{3}}(\bar{p}_{2})\tilde{J}^{d\,
\lambda_{4}}(\bar{p}_{3})\rrangle\right]}{4\text{Re}[(-i)^{d}\llangle T(\bar{p}_{1})T(-\bar{p}_{1})\rrangle]
\text{Re}[(-i)^{d}\llangle\tilde{J}^{\rho}_{a}(\bar{p}_{2})\tilde{J}^{\lambda_{3}}_{c}(-\bar{p}_{2})\rrangle]
\text{Re}[(-i)^{d}\llangle\tilde{J}^{\sigma}_{b}(\bar{p}_{3})\tilde{J}^{\lambda_{4}}_{d}(-\bar{p}_{3})\rrangle]} ,
\label{holographic_formula_xiAA}
\end{align}
where from now on we leave implicit that the analytical continuation of the correlators is performed 
\emph{before} taking their real part, for the sake of notation. We can use the holographic formulas for the 
2-point function \eqref{holographic_formula_2pt_function} to rewrite \eqref{holographic_formula_xiAA} in a 
way that the cosmological correlators are explicit:
\begin{align}
    \llangle\xi(p_{1})\tilde{A}^{a}_{\rho}(p_{2})\tilde{A}^{b}_{\sigma}(p_{3})\rrangle & = -2\mathrm{Re}\left[(-
    i)^{d}\llangle T(\bar{p}_{1})\tilde{J}^{\lambda_{3}}_{c}(\bar{p}_{2})\tilde{J}^{\lambda_{4}}_{d}(\bar{p}_{3})
    \rrangle\right] \nonumber \\
    & \qquad \times \llangle\xi(p_{1})\xi(-p_{1})\rrangle\llangle\tilde{A}^{a}_{\rho}(p_{2})\tilde{A}^{c}
    _{\lambda_{3}}(-p_{2})\rrangle\llangle\tilde{A}^{b}_{\sigma}(p_{3})\tilde{A}^{d}_{\lambda_{4}}(-p_{3})
    \rrangle. \label{definition_xi_A_A}
\end{align}
Let us define the dimensionful power spectrum $\mathcal{P}_{\tilde{A}}(p)$ associated with the monopole 
field and $\mathcal{P}_\xi(p)$ associated with the scalar curvature as
\begin{equation}
    \llangle\tilde{A}^{a}_{\mu}(p)\tilde{A}^{b}_{\nu}(-p)\rrangle\equiv\mathcal{P}_{\tilde{A}}(p)\delta^{ab}
    \pi_{\mu\nu} \qquad \text{and } \qquad \llangle\xi(p_{1})\xi(-p_{1})\rrangle\equiv\mathcal{P}_{\xi}
    (p_{1}) \;.
\end{equation}
With these definitions, we can write the formula \eqref{definition_xi_A_A} schematically as 
\begin{align}
    & \llangle\xi(p_{1})\tilde{A}^{a}_{\rho}(p_{2})\tilde{A}^{b}_{\sigma}(p_{3})\rrangle = \nonumber \\
    & =-2\mathrm{Re}\left[(-i)^{d}\llangle T(\bar{p}_{1})\tilde{J}^{a\lambda_{3}}(\bar{p}_{2})\tilde{J}
    ^{b\lambda_{4}}(\bar{p}_{3})\rrangle\right]\pi_{\rho\lambda_{3}}(p_{2})\pi_{\sigma\lambda_{4}}(p_{3})
    \mathcal{P}_{\xi}(p_{1})\mathcal{P}_{\tilde{A}}(p_{2})\mathcal{P}_{\tilde{A}}(p_{3})\, . 
    \label{main_holo_formula_xi_AA}
\end{align}

The amplitude of the vortex power spectrum was calculated holographically using the electric Noether 
current $J_\mu^a$ within our toy model in \cite{nastase_pre-inflationary}, using the formula 
\begin{equation}
    \llangle\tilde{A}^{a}_{\mu}(p)\tilde{A}^{b}_{\nu}(-p)\rrangle=-\frac{1}{2\text{Re}[(-i)^{d}\llangle\tilde{J}
    ^{\mu}_{a}(\bar{p})\tilde{J}^{\nu}_{b}(-\bar{p})\rrangle]},\label{holo_formula_AA}
\end{equation}
where the magnetic vortex current 2-point function is obtained from the electric correlator by 
$Sl(2,\mathbb{Z})$ duality. Explicitly, we parametrize the 2-point function of the electric current as 
\eqref{2pt_function_JJ_t_omega}, and use the action of $Sl(2,\mathbb{Z})$ duality, which is to invert the 
constant $t\rightarrow t^{-1}$. This gives, after analytical continuation and inverting the resulting operator,
\begin{equation}
    \llangle\tilde{A}^{a}_{\mu}(p)\tilde{A}^{b}_{\nu}(-p)\rrangle = -\frac{1}{2\text{Re}[(-i)^{d}\llangle\tilde{J}
    ^{\mu}_{a}(\bar{p})\tilde{J}^{\nu}_{b}(-\bar{p})\rrangle]} = -\frac{\pi^{2}N^{2}}{2p}\delta_{ab}\pi^{\mu\nu},
\end{equation}
from which we obtain 
\begin{equation}
    \mathcal{P}_{\tilde{A}}(p)=-\frac{\pi^{2}N^{2}}{2p}.
\end{equation}

Note that the transverse projector $\pi_{\mu \nu}$ in \eqref{holo_formula_AA} is not invertible in the usual 
sense, since it has zero eigenvalues. Here we use the definition of \emph{Moore–Penrose 
pseudo-inverse}, discussed in more detail in \autoref{append_inversion}. Formally, for a given matrix 
$A$, its 
pseudo inverse is defined by the formula $A A^+ A = A$ (besides other conditions), and from the 
properties of the transverse projector (symmetric and idempotency), one can verify that $\pi_{\mu\alpha} 
\pi^{\alpha\beta} \pi_{\beta\nu} = \pi_{\mu\nu}$. We conclude that the transverse projector is its own 
pseudoinverse, with indices reallocated accordingly, and one can check that the same is true for the 
transverse-traceless projector.

For the scalar power spectrum, we use again the holographic formula 
\eqref{holographic_formulas_scalar_tensor_power_spectrum} to write 
\begin{equation}
    \mathcal{P}_{\xi}(p_{1}) = -\frac{1}{8}\frac{1}{\text{Re}[(-i)^{d}\bar{B}(\bar{p})]},
\end{equation}
where $\bar B(\bar p)$ is defined by the decomposition \eqref{tensor_decomposition_2pt_function_TT}. 
The general result for the class of phenomenological models considered in \cite{holouniverse} is given in 
terms of the field content by
\begin{equation}
    \bar{A}(\bar{p})=\frac{\bar{N}^{2}}{256}\bar{p}^{3}\left(\mathcal{N}_{A}+\mathcal{N}_{\Phi}+\mathcal{N}
    _{\chi}+2\mathcal{N}_{\psi}\right),\qquad\qquad\bar{B}(\bar{p})=\frac{\bar{N}^{2}}{256}\bar{p}
    ^{3}\left(\mathcal{N}_{A}+\mathcal{N}_{\Phi}\right),
\end{equation}
and applying to our toy model, we set $\mathcal{N}_{A}=1$, $\mathcal{N}_{\Phi}=12$, and $\mathcal{N}
_{\chi}=0=\mathcal{N}_{\psi}$ (note that we have two types of complex scalar $SO(3)$ triplets, so $12$ 
scalar degrees of freedom in total), and we find that both coefficients are equal in our toy model, such 
that, after analytical continuation, we find 
\begin{equation}
    \text{Re}[(-i)^{d}\bar{A}(\bar{p}_{1})]=\text{Re}[(-i)^{d}\bar{B}(\bar{p}_{1})]= -\frac{13}{256}N^{2}p^{3}
    _{1}\,.
\end{equation}
which means that at 1-loop we have 
\begin{equation}
    \mathcal{P}_{\xi}(p_{1}) = -\frac{1}{8} \left( -\frac{13}{256} N^2 p_1^3 \right)^{-1} = \frac{32}{13}\frac{1}
    {N^{2}p^{3}_{1}}.
\end{equation}

Note that the (dimensionless) power spectrum is usually defined by taking out a $2\pi^2/p_1^3$, so 
\be
\tilde P_\xi(p_1)=\frac{32}{13}\frac{1}{2\pi^2N^2}\;,
\ee
which led \cite{planckdata} to fix $N\simeq 3000$ by matching with CMBR observation (combined 
with the matching of the full spectrum). If we also define a similar dimensionless power spectrum for $\tilde A$, 
\be
\tilde P_{\tilde A}(p)\equiv\frac{p}{2\pi^2}P_{\tilde A}(p)=-\frac{N^2}{4}\;,
\ee
with the value of $N$ from $\tilde P_\xi$ we would get $\tilde P_{\tilde A}\sim 2\times 10^6$, though 
this could be interpreted as possibly due to the dimensionless size squared of the monopoles.

We can do the same derivation for the cross correlator with the tensor perturbation $\gamma_{\mu \nu}$. 
We start with the basic expression
\begin{align}
    &\llangle\gamma_{\mu\nu}(p_{1})\tilde{A}^{a}_{\rho}(p_{2})\tilde{A}^{b}_{\sigma}(p_{3})\rrangle  
    =\nonumber \\ 
    &  \frac{\mathrm{\text{Re}}\left[(-i)^{d}\llangle T^{\lambda_{1}\lambda_{2}}_{\text{TT}}(\bar{p}_{1})
    \tilde{J}^{\lambda_{3}}_{c}(\bar{p}_{2})\tilde{J}^{\lambda_{4}}_{d}(\bar{p}_{3})\rrangle\right] }{\text{Re}[(-
    i)^{d}\llangle T^{\mu\nu}_{\text{TT}}(\bar{p}_{1})T^{\lambda_{1}\lambda_{2}}_{\text{TT}}(-\bar{p}_{1})
    \rrangle]\text{Re}[(-i)^{d}\llangle\tilde{J}^{\rho}_{a}(\bar{p}_{2})\tilde{J}^{\lambda_{3}}_{c}(-\bar{p}_{2})
    \rrangle]\text{Re}[(-i)^{d}\llangle\tilde{J}^{\sigma}_{b}(\bar{p}_{3})\tilde{J}^{\lambda_{4}}_{d}(-\bar{p}_{3})
    \rrangle]} ,
\end{align}
which can be written in terms of the 2-point cosmological observables using the previous definitions:
\begin{align}
    \llangle\gamma_{\mu\nu}(p_{1})\tilde{A}^{a}_{\rho}(p_{2})\tilde{A}^{b}_{\sigma}(p_{3})\rrangle & 
    = -\frac{1}{2}\mathrm{\text{Re}}\left[(-i)^{d}\llangle T^{\lambda_{1}\lambda_{2}}_{\text{TT}}(\bar{p}_{1})
    \tilde{J}^{a\lambda_{3}}(\bar{p}_{2})\tilde{J}^{d\lambda_{4}}(\bar{p}_{3})\rrangle\right] \nonumber \\
    & \quad \times \pi_{\rho\lambda_{3}}(p_{2})\pi_{\sigma\lambda_{4}}(p_{3})\llangle\gamma_{\mu\nu}
    (p_{1})\gamma_{\lambda_{1}\lambda_{2}}(-p_{1})\rrangle\mathcal{P}_{\tilde{A}}(p_{2})\mathcal{P}
    _{\tilde{A}}(p_{3}). \label{gammaAA_holographic}
\end{align}

Using the definition of the transverse-traceless projection, we have 
\begin{equation}
   \llangle T^{\lambda_{1}\lambda_{2}}_{\text{TT}}(\bar{p}_{1})\tilde{J}^{a\lambda_{3}}(\bar{p}_{2})\tilde{J}
   ^{d\lambda_{4}}(\bar{p}_{3})\rrangle = \llangle T^{\alpha\beta}(\bar{p}_{1})\tilde{J}^{a\lambda_{3}}
   (\bar{p}_{2})\tilde{J}^{d\lambda_{4}}(\bar{p}_{3})\rrangle\Pi^{\lambda_{1}\lambda_{2}}{}_{\alpha\beta}
   (\bar p_{1}),
\end{equation}
such that we can act with the projector $\Pi^{\lambda_{1}\lambda_{2}}{}_{\alpha\beta}(\bar p_{1})$ to 
simplify the result \eqref{gammaAA_holographic}:
\begin{equation}
    \Pi^{\lambda_{1}\lambda_{2}}{}_{\alpha\beta}(p_{1})\llangle\gamma_{\mu\nu}(p_{1})
    \gamma_{\lambda_{1}\lambda_{2}}(-p_{1})\rrangle=-\frac{8\Pi^{\lambda_{1}\lambda_{2}}{}
    _{\alpha\beta}(p_{1})}{\text{Re}[(-i)^{d}\llangle T^{\mu\nu}_{TT}(\bar{p}_{1})T^{\lambda_{1}\lambda_{2}}
    _{TT}(-\bar{p}_{1})\rrangle]}.
\end{equation}
For the correlator in the denominator, we use the result
\begin{align}
    \llangle T^{\mu\nu}_{TT}(\bar{p}_{1})T^{\lambda_{1}\lambda_{2}}_{TT}(-\bar{p}_{1})\rrangle & = 
    \Pi^{\mu\nu}{}_{\rho\sigma}(\bar{p}_{1})\Pi^{\lambda_{1}\lambda_{2}}{}_{\gamma\delta}(\bar{p}_{1})
    \llangle T^{\rho\sigma}(\bar{p}_{1})T^{\gamma\delta}(-\bar{p}_{1})\rrangle \nonumber \\
    & = \bar{A}(\bar{p}_{1})\Pi^{\mu\nu\lambda_{1}\lambda_{2}}(p_{1}),
\end{align}
such that
\begin{align}
    \Pi^{\lambda_{1}\lambda_{2}}{}_{\alpha\beta}(p_{1})\llangle\gamma_{\mu\nu}(p_{1})
    \gamma_{\lambda_{1}\lambda_{2}}(-p_{1})\rrangle&=-\frac{8\Pi^{\lambda_{1}\lambda_{2}}{}
    _{\alpha\beta}(p_{1})\Pi_{\mu\nu\lambda_{1}\lambda_{2}}(p_{1})}{\text{Re}[(-i)^{d}\bar{A}(\bar{p}_{1})]} 
    \nonumber \\
    & = -\frac{8\Pi_{\mu\nu\alpha\beta}(p_{1})}{\text{Re}[(-i)^{d}\bar{A}(\bar{p}_{1})]},
\end{align}
where again, we use the idempotency of the transverse-traceless projector to define its pseudo-inverse as 
$(\Pi_{\mu \nu \rho \sigma})^{-1} = \Pi^{\mu \nu \rho \sigma}$. Following the conventions we used for the 
scalar perturbations, we define the dimensionful power spectrum for the tensor perturbations as 
\begin{equation}
    \mathcal{P}_{\gamma}(p)\equiv-\frac{8}{\text{Re}[(-i)^{d}\bar{A}(\bar{p})]},
\end{equation}
and the holographic formula for the cosmological correlator $\langle \gamma \tilde{A} \tilde{A} \rangle$ 
becomes 
\begin{align}
    \llangle\gamma_{\mu\nu}(p_{1})\tilde{A}^{a}_{\rho}(p_{2})\tilde{A}^{b}_{\sigma}(p_{3})\rrangle & =-
    \frac{1}{2}\mathrm{\text{Re}}\left[(-i)^{d}\llangle T^{\alpha\beta}(\bar{p}_{1})\tilde{J}^{a\lambda_{3}}
    (\bar{p}_{2})\tilde{J}^{d\lambda_{4}}(\bar{p}_{3})\rrangle\right] \nonumber \\ 
    & \qquad \times \pi_{\rho\lambda_{3}}(p_{2})\pi_{\sigma\lambda_{4}}(p_{3})\Pi_{\mu\nu\alpha\beta}
    (p_{1})\mathcal{P}_{\gamma}(p_{1})\mathcal{P}_{\tilde{A}}(p_{2})\mathcal{P}_{\tilde{A}}(p_{3}). 
    \label{main_holo_formula_gamma_AA}
    \end{align}

Finally, the last ingredient we need to compute the cosmological correlators is the QFT correlator $\langle 
TJJ\rangle$, which is the main goal of the rest of this paper.
%
%
%
%
%

\section{Calculation of 3-point function $\la TJJ \ra$}\label{sec:calculation}

When dealing with 3-point functions involving the energy-momentum tensor, one must be careful about 
contact terms, which arise when the insertion points of the operator coincide. The energy-momentum 
tensor is defined through the functional derivative of the action with respect to the background metric,
\begin{equation}
  T^{\mu \nu}(x)=\frac{2}{\sqrt{g}} \left.\frac{\delta S}{\delta g_{\mu \nu}(x)} \right|_{g_{\mu\nu}
  =\delta_{\mu\nu}}.
\end{equation}
 To compute 3-point functions one takes three functional derivatives of the generating functional 
 $Z[g_{\mu \nu},a_\mu]$ with respect to the source. In our case,
\begin{equation}\label{TJJfulldef}
  \left\langle T_{\mu\nu}(x_{1})J_{\rho}(x_{2})J_{\sigma}(x_{3})\right\rangle = \frac{-1}{\sqrt{g(x_{3})}}
  \frac{\delta}{\delta a^{\sigma}(x_{3})}  \frac{-1}{\sqrt{g(x_{2})}}\frac{\delta}{\delta a^{\rho}(x_{2})} \frac{-2}
  {\sqrt{g(x_{1})}}\frac{\delta}{\delta g^{\mu\nu}(x_{1})} Z[g_{\mu \nu},a_\mu],
\end{equation}
where $a_\mu$ is the background vector field that sources the $SO(3)$ global current $J_{\mu}$ and 
$g_{\mu\nu}$ is the 3-metric that sources the stress-energy tensor $T_{\mu\nu}$. 

Note that, even if we are dealing with a QFT in flat space, we must not take the sources to zero $g 
\rightarrow 0,a \rightarrow 0$ until all the derivatives are taken. As we show in \autoref{appendA}, this 
gives rise to extra terms that sum up to the usual QFT concept of a 3-point function. In momentum 
space, the full three-point correlator is 
\begin{align}\label{CTP}
    &\llangle T_{\mu\nu}(\bar{p}_{1})J_{\rho}(\bar{p}_{2})J_{\sigma}(\bar{p}_{3})\rrangle_{\text{full}} 
     \notag\\
    &=\llangle T_{\mu\nu}(\bar{p}_{1})J_{\rho}(\bar{p}_{2})J_{\sigma}(\bar{p}_{3})
    \rrangle_{\text{flat}}-2\delta_{\rho(\mu}\llangle J_{\nu)}(-\bar{p}_{3})J_{\sigma}(\bar{p}_{3})
    \rrangle-2\delta_{\sigma(\mu}\llangle J_{\nu)}(-\bar{p}_{2})J_{\rho}(\bar{p}_{2})\rrangle \notag\\
    & \hspace{1cm} +\delta_{\mu\nu}\left(\llangle J_{\rho}(-\bar{p}_{3})J_{\sigma}(\bar{p}_{3})\rrangle+
    \llangle J_{\sigma}(-\bar{p}_{2})J_{\rho}(\bar{p}_{2})\rrangle\right),
\end{align}
where 
\begin{equation}\label{TJJflatdef}
  \llangle T_{\mu\nu}(\bar{p}_{1}) J_{\rho}(\bar{p}_{2})J_{\sigma}(\bar{p}_{3})\rrangle_{\text{flat}} = 
  \int\mathcal{D}\Phi e^{-S_{\text{tot}}} T_{\mu\nu}(\bar{p}_{1})J_{\rho}(\bar{p}_{2})J_{\sigma}(\bar{p}_{3}),
\end{equation}
is the correlator obtained by three separate insertions of the operators and it accounts only for the physics 
in non-coincident points. This ``flat" correlator is the one we compute using Feynman diagrams, so in the 
end, we must add the contact terms in \eqref{CTP} to get the full correlator. 

Now we compute the $\langle TJJ \ra$ correlator following the toy model \eqref{toymodel} proposed in 
\cite{nastase_pre-inflationary}. This 3-point function is composed by one insertion of the energy-
momentum tensor and two insertions of the global current conserved in the model. At 1-loop order, we 
have only one diagram 
\begin{equation}
    \llangle T_{\mu\nu}(\bar{p}_1) J^a_{\rho}(-\bar{p}_2) J^b_{\sigma}(-\bar{p}_3) \rrangle_{\text{flat} } = 
    \begin{tikzpicture}[baseline = (a.base),arrowlabel/.style={
        /tikzfeynman/momentum/.cd, 
        arrow shorten=#1,arrow distance=2.5mm
      },
      arrowlabel/.default=0.4]
        \def\leglength{1}
        \begin{feynman} [inline =(a.base) ]
            \vertex[ crossed dot, label = left:{$\mu,\nu$}] (a) at (-\leglength,0) {};
            \vertex[crossed dot, label = right:{$a,\rho$}]  (b) at ( \leglength,\leglength) {};
            \vertex[crossed dot, label = right:{$b,\sigma$}]  (c) at (\leglength,- \leglength) {};
    
    \diagram*{
        (a) -- [fermion,momentum ={[arrowlabel]$q+\bar{p}_2$}] (b) -- [fermion, momentum={[arrowlabel]$q$}] 
        (c) -- [fermion, momentum={[arrowlabel]$q- \bar{p}_3$}] (a),
        (l1) [momentum ={[arrowlabel]$q+\bar{p}_2$}] (a),
    };
        \end{feynman}
    \end{tikzpicture},
\end{equation}
in which the momentum $\bar{p}_1$ flows into the diagram from the left and $\bar{p}_2+\bar{p}_3$ flows 
out to the right (hence the minus sign for the momentum of the current insertions).

After applying the Feynman rules, there are different ways we can write the integral expression for the 
triangle diagram, depending on our choice of independent momenta. We choose to write everything in 
terms of the vectors associated with momenta $\bar{p}_2$ and $\bar{p}_3$. By doing so, the correlator 
can be decomposed into its Lorentz structures and a set of scalar integrals:
\begin{align}\label{correlator}
    \llangle T_{\mu\nu}(\bar{p}_{1})J^{a}_{\rho}(\bar{p}_{2})J^{b}_{\sigma}(\bar{p}_{3})\rrangle_{\text{flat}} & 
    = 8 \bar{N}^{2}\delta^{ab}\mathcal{F}_{\mu\nu\rho\sigma}(\bar{p}_{1},\bar{p}_{2},\bar{p}_{3}),
\end{align}
where 
\begin{align}
    \mathcal{F}_{\mu\nu\rho\sigma}(\bar{p}_{1},\bar{p}_{2},\bar{p}_{3}) & =- 8 S_{\mu\rho\sigma\nu} - 
    \mathcal{\bar K}_{\mu\nu}\left(\frac{1}{2}\bar{p}_{2\rho}\bar{p}_{3\sigma}S+\bar{p}_{3\sigma}S_{\rho}-
    \bar{p}_{2\rho}S_{\sigma}-2S_{\rho\sigma}\right)  \nonumber \\
    & \hspace{-2.5cm} - 4\left[ (\bar{p}_{2} - \bar{p}_{3})_{\nu} S_{\mu\rho\sigma} + (\bar{p}_{2} - \bar{p}
    _{3})_{\mu}S_{\rho\sigma\nu} + \bar{p}_{2\rho} S_{\mu\sigma\nu} - \bar{p}_{3\sigma}S_{\mu\rho\nu} - 
    \delta_{\mu\nu}(\bar{p}_{2} - \bar{p}_{3})^{\eta}S_{\eta\rho\sigma}\right] \nonumber \\
    & \hspace{-2.5cm} + \mathcal{\bar J}_{\nu\sigma}S_{\mu\rho} - \mathcal{\bar P}_{\nu\rho}
    S_{\mu\sigma} + \mathcal{\bar J}_{\mu\sigma}S_{\rho\nu} - \mathcal{\bar P}_{\mu\rho}S_{\sigma\nu} - 
    \delta_{\mu\nu}\mathcal{\bar J}^{\eta}_{\,\,\sigma}S_{\eta\rho}+\delta_{\mu\nu}\mathcal{\bar P}^{\eta}
    _{\,\,\rho}S_{\eta\sigma} \nonumber \\
    & \hspace{-2.5cm} - 2 \delta_{\mu\nu}(\bar{p}_{3\sigma}T_{\rho} - \bar{p}_{2\rho}T_{\sigma})- 4 
    \delta_{\mu\nu}T_{\rho\sigma} - \bar{p}_{2\rho}\bar{p}_{3\sigma}\left(\delta_{\mu\nu}T-2S_{\mu\nu}
    \right)\nonumber \\
    & \hspace{-2.5cm} +\frac{1}{2} \left(\bar{p}_{2\rho}\mathcal{\bar J}_{\nu\sigma}S_{\mu}+\bar{p}_{2\rho}
    \mathcal{\bar J}_{\mu\sigma}S_{\nu}-\delta_{\mu\nu}\bar{p}_{2\rho}\mathcal{\bar J}^{\eta}_{\,\,\sigma}
    S_{\eta}\right), 
\end{align}
and we have defined the following Lorentz structures that depend only on the external momenta $\bar 
p_2$ and $\bar p_3$
\begin{align}
    \mathcal{\bar K}_{\mu\nu} & = 2(\bar{p}_{2\mu}\bar{p}_{3\nu}+\bar{p}_{2\nu}\bar{p}_{3\mu}-g_{\mu\nu}
    (\bar{p}_{2}\cdot\bar{p}_{3})), \nonumber \\
    \mathcal{\bar J}_{\mu\nu} & = 2(\bar{p}_{2\mu}\bar{p}_{3\nu}-\bar{p}_{3\mu}\bar{p}_{3\nu}), \nonumber 
    \\
    \mathcal{\bar P}_{\mu\nu} & = 2(\bar{p}_{2\mu}\bar{p}_{2\nu}-\bar{p}_{2\nu}\bar{p}_{3\mu}),
\end{align}
to simplify the notation. The  pre-factor  $ 8 \bar{N}^{2}\delta^{ab}$ comes from the symmetry of the 
diagram and the contraction of the scalar indices $(i,j,...)$, the $SO(3)$ indices and the implicit 
$SU(\bar{N})$ indices. Moreover, the $S_{\mu,\nu,...}$ and $T_{\mu,\nu,...}$ terms are integrals over the 
loop momentum $q$, and they are defined as
\begin{align}\label{SandTdef}
S &\equiv \int\!\frac{d^{d}q}{(2\pi)^{d}}\frac{1}{q^{2}(q+\bar{p}_{2})^{2}(q-\bar{p}_{3})^{2}}\,,
& \qquad
T &\equiv \int\!\frac{d^{d}q}{(2\pi)^{d}}\frac{1}{(q+\bar{p}_{2})^{2}(q-\bar{p}_{3})^{2}}\,,
\\
S_{\mu} &\equiv \int\!\frac{d^{d}q}{(2\pi)^{d}}\frac{q_{\mu}}{q^{2}(q+\bar{p}_{2})^{2}(q-\bar{p}_{3})^{2}}\,,
& \qquad
T_{\mu} &\equiv \int\!\frac{d^{d}q}{(2\pi)^{d}}\frac{q_{\mu}}{(q+\bar{p}_{2})^{2}(q-\bar{p}_{3})^{2}}\,,
\\
S_{\mu\nu} &\equiv \int\!\frac{d^{d}q}{(2\pi)^{d}}\frac{q_{\mu}q_{\nu}}{q^{2}(q+\bar{p}_{2})^{2}(q-\bar{p}
_{3})^{2}}\,,
& \qquad
T_{\mu\nu} &\equiv \int\!\frac{d^{d}q}{(2\pi)^{d}}\frac{q_{\mu}q_{\nu}}{(q+\bar{p}_{2})^{2}(q-\bar{p}_{3})^{2}}
\,,
\\
S_{\mu\nu\rho} &\equiv \int\!\frac{d^{d}q}{(2\pi)^{d}}\frac{q_{\mu}q_{\nu}q_{\rho}}{q^{2}(q+\bar{p}_{2})^{2}
(q-\bar{p}_{3})^{2}}\,,
& \qquad
S_{\mu\nu\rho\sigma} &\equiv \int\!\frac{d^{d}q}{(2\pi)^{d}}\frac{q_{\mu}q_{\nu}q_{\rho}q_{\sigma}}{q^{2}
(q+\bar{p}_{2})^{2}(q-\bar{p}_{3})^{2}}\,.
\end{align}

We can decompose these integrals in terms of Lorentz tensors using the Feynman parameterization as 
discussed in \autoref{appendB}.
Finally, computing the correlator $\langle TJJ \rangle$  reduces to solving all remaining scalar integrals 
that appear in the $S$ and $T$ tensor decompositions. All of them are finite, both in the UV and in the IR 
--- the details of the calculation is left in the \autoref{appendB}, and the results are listed in 
\autoref{append_list}. 
With those in hands, we can replace them in the definitions of the tensorial integrals $S$ and, together 
with the results for $T$, find the final expression for the 3-point function \eqref{correlator}. After adding the 
contact terms \eqref{CTP}, we write the result for the full correlator as

\begin{equation}\label{fullcorrelator}
\begin{split}
\llangle T_{\mu\nu}(\bar{p}_{1})J^{a}_{\rho}(\bar{p}_{2})J^{b}_{\sigma}(\bar{p}_{3})\rrangle_{\text{full}} &=  
\bar{N}^2\delta^{ab} \big[ A_{(1)} \delta_{\mu \nu} \delta_{\rho \sigma} + A_{(2)} (\delta_{\mu 
\sigma} \delta_{\rho \nu} + \delta_{\mu \rho} \delta_{\sigma \nu}) \\
&\hspace{-4cm}+ A_{(3)} (\delta_{\sigma \nu} \bar{p}_{2\mu} \bar{p}_{2\rho} + \delta_{\mu 
\sigma} \bar{p}_{2\rho} \bar{p}_{2\nu}) + A_{(4)} (\delta_{\rho \nu} \bar{p}_{2\mu} \bar{p}
_{2\sigma} + \delta_{\mu \rho} \bar{p}_{2\sigma} \bar{p}_{2\nu})  \\
&\hspace{-4cm}+ A_{(5)} \delta_{\mu \nu} \bar{p}_{2\rho} \bar{p}_{2\sigma} + A_{(6)} \delta_{\rho 
\sigma} \bar{p}_{2\mu} \bar{p}_{2\nu} + A_{(7)} \bar{p}_{2\mu} \bar{p}_{2\rho} \bar{p}_{2\sigma} 
\bar{p}_{2\nu}  \\
& \hspace{-4cm} + A_{(8)} (\delta_{\sigma \nu} \bar{p}_{2\rho} \bar{p}_{3\mu} + \delta_{\rho 
\nu} \bar{p}_{2\mu} \bar{p}_{3\sigma} + \delta_{\mu \rho} \bar{p}_{2\nu} \bar{p}_{3\sigma} + 
\delta_{\mu \sigma} \bar{p}_{2\rho} \bar{p}_{3\nu}) \\
&\hspace{-4cm}+ A_{(9)} (\delta_{\rho \nu} \bar{p}_{2\sigma} \bar{p}_{3\mu} + \delta_{\mu 
\rho} \bar{p}_{2\sigma} \bar{p}_{3\nu}) + A_{(10)} (\delta_{\rho \sigma} \bar{p}_{2\nu} \bar{p}
_{3\mu} + \delta_{\rho \sigma} \bar{p}_{2\mu} \bar{p}_{3\nu})  \\
&\hspace{-4cm} + A_{(11)} (\bar{p}_{2\rho} \bar{p}_{2\sigma} \bar{p}_{2\nu} \bar{p}_{3\mu} + 
\bar{p}_{2\mu} \bar{p}_{2\rho} \bar{p}_{2\sigma} \bar{p}_{3\nu}) \\
&\hspace{-4cm}+ A_{(12)} (\delta_{\sigma \nu} \bar{p}_{2\mu} \bar{p}_{3\rho} + \delta_{\mu 
\sigma} \bar{p}_{2\nu} \bar{p}_{3\rho}) + A_{(13)} \delta_{\mu \nu} \bar{p}_{2\sigma} \bar{p}
_{3\rho} + A_{(14)} \bar{p}_{2\mu} \bar{p}_{2\sigma} \bar{p}_{2\nu} \bar{p}_{3\rho} \\
&\hspace{-4cm}+ A_{(15)} (\delta_{\sigma \nu} \bar{p}_{3\mu} \bar{p}_{3\rho} + \delta_{\mu 
\sigma} \bar{p}_{3\rho} \bar{p}_{3\nu}) + A_{(16)} (\bar{p}_{2\sigma} \bar{p}_{2\nu} \bar{p}
_{3\mu} \bar{p}_{3\rho} + \bar{p}_{2\mu} \bar{p}_{2\sigma} \bar{p}_{3\rho} \bar{p}_{3\nu}) \\
&\hspace{-4cm} + A_{(17)} \delta_{\mu \nu} \bar{p}_{2\rho} \bar{p}_{3\sigma}  + A_{(18)} \bar{p}
_{2\mu} \bar{p}_{2\rho} \bar{p}_{2\nu} \bar{p}_{3\sigma} + A_{(19)} (\delta_{\rho \nu} \bar{p}
_{3\mu} \bar{p}_{3\sigma} + \delta_{\mu \rho} \bar{p}_{3\sigma} \bar{p}_{3\nu})  \\
&\hspace{-4cm} + A_{(20)} (\bar{p}_{2\rho} \bar{p}_{2\nu} \bar{p}_{3\mu} \bar{p}_{3\sigma} + 
\bar{p}_{2\mu} \bar{p}_{2\rho} \bar{p}_{3\sigma} \bar{p}_{3\nu}) + A_{(21)} \delta_{\mu \nu} 
\bar{p}_{3\rho} \bar{p}_{3\sigma}  \\
&\hspace{-4cm}+ A_{(22)} \bar{p}_{2\mu} \bar{p}_{2\nu} \bar{p}_{3\rho} \bar{p}_{3\sigma} + 
A_{(23)} (\bar{p}_{2\nu} \bar{p}_{3\mu} \bar{p}_{3\rho} \bar{p}_{3\sigma} + \bar{p}_{2\mu} \bar{p}
_{3\rho} \bar{p}_{3\sigma} \bar{p}_{3\nu}) \\
&\hspace{-4cm}+ A_{(24)} \delta_{\rho \sigma} \bar{p}_{3\mu} \bar{p}_{3\nu} + A_{(25)} \bar{p}
_{2\rho} \bar{p}_{2\sigma} \bar{p}_{3\mu} \bar{p}_{3\nu} + A_{(26)} \bar{p}_{2\sigma} \bar{p}
_{3\mu} \bar{p}_{3\rho} \bar{p}_{3\nu} \\
&\hspace{-4cm}+ A_{(27)} \bar{p}_{2\rho} \bar{p}_{3\mu} \bar{p}_{3\sigma} \bar{p}_{3\nu} + 
A_{(28)} \bar{p}_{3\mu} \bar{p}_{3\rho} \bar{p}_{3\sigma} \bar{p}_{3\nu} \big],
\end{split}
\end{equation}
where the form factors $ A_{(i)}(\bar{p}_1,\bar{p},\bar{p}_3)$ are listed in \autoref{append_formfactors}.

This is the three-point function involving the stress-energy tensor and two insertions of a SO(3) global 
current of a quantum field theory dual to a domain-wall spacetime. Before we move on to analyze this 
result, we can verify it using Ward identities. In particular, the energy-momentum tensor is the conserved 
current
under diffeomorphisms, and it is required that the three-point function respects this symmetry.
We show the derivation of the transverse Ward identity in \autoref{appendC}, and it amounts to
\begin{equation}\label{WItransv}
    p_{1}^{\mu} \llangle T_{\mu\nu}(p_{1})J_{\rho}(p_{2})J_{\sigma}(p_{3}) \rrangle _{\text{full}} = 
    {2\delta^{[\alpha}_{\rho}p_{2\nu]} \llangle J_{\alpha}(-p_{3})J_{\sigma}(p_{3}) \rrangle +2\delta^{[\alpha}
    _{\sigma}p_{3\nu]} \llangle J_{\mu}(-p_{2})J_{\rho}(p_{2}) \rrangle } ,
\end{equation}
Indeed, contracting our result \eqref{fullcorrelator} with $p_{1}^{\mu}$, we find consistency with the 
constraint \eqref{WItransv}. To check that, we used the result of the 2-point function 
\begin{equation}
  \llangle J^a_{\mu}(\bar{p})J^b_{\nu}(-\bar{p})\rrangle = \bar{N}^2 \frac{\bar{p}}{4} \delta^{ab} 
  \left(\delta_{\mu\nu}-\frac{\bar{p}_{\mu}\bar{p}_{\nu}}{\bar{p}^{2}}\right)\, ,
\end{equation}
derived in \cite{nastase_pre-inflationary}.

Aditionally to the Ward identity related to the conservation of the stress-energy tensor, one can also expect the correlator to respect the conservation of the vortex current $J$. In fact, in \cite{3ptward} it is shown that one can decompose such correlators into a term that is transverse (or transverse-traceless, for a CFT) plus a term that can be reconstructed entirely from these Ward identities.

Whereas we are not exploring this method here, we can still extract, by completeness, the transverse part of $\langle TJJ \rangle$ to find that it can be written in terms of 7 independent tensor structures. This is the same number of structures one would find by following the procedure outlined in \cite{3ptward}, even though we adopt a different basis and momentum convention. The steps to obtain the projected result are detailed in \autoref{appendA}.
\subsection{Soft limit}

After displaying the general result for the correlator $\left\langle TJJ\right\rangle$, we would like to 
examine the soft limit, as discussed in \autoref{cosmonongauss}. On the cosmology side, we want to 
consider the scalar $\xi(p_1)$ and tensor $\gamma_{\mu \nu} (p_1)$ perturbations as the soft modes, 
which means taking to zero the momentum associated with the insertion of the energy-momentum tensor 
in the QFT dual correlator.

Since the limit $p_1\rightarrow 0$ may be divergent, we can regulate the squeezed limit by considering 
the following parameterization for the momenta \footnote{The above parameterization was introduced in 
the context of investigating IR divergences for 3-point functions in the squeezed limit for mass-deformed 
theories\cite{matheusIRdivergences}.}
\begin{equation}
    \bar{p}_{1\mu}=2\varepsilon \bar{p}_{\mu},\qquad \bar{p}_{2\mu}=(1+\varepsilon)\bar{p}_{\mu}, \qquad 
    \bar{p}_{3\mu}=-(1-\varepsilon)\bar{p}_{\mu}.\label{def_squeezed_limit}
\end{equation}
This choice satisfies $\bar{p}_{1\mu}=\bar{p}_{2\mu}+\bar{p}_{3\mu}$. To apply the soft limit, we take $
\varepsilon\ll1$, which results in $\bar{p}_{1}^{2}\ll \bar{p}_{2}^{2},\bar{p}_{3}^{2}$ and $\bar{p}_{2\mu}
\simeq -\bar{p}_{3\mu}=\bar{p}_{\mu}$. Note that in this choice of parametrizations we are taking the 
squeezed limit by placing the three vectors in a particular collinear shape from the start. This is not the 
most generic choice, but it is enough for the cases where the observables depend mostly on the 
magnitudde of the momenta instead on the specific direction of the momentum vector in space. 

To find the soft limit for the 3-point function, we apply this parameterization to the form factors obtained in 
the general result \eqref{fullcorrelator}, by replacing the absolute value of the momenta $\bar{p}_{1}
=2\varepsilon \bar{p}$, $\bar{p}_{2}=\left(1+\varepsilon\right)\bar{p}$ and $\bar{p}_{3}=(1-\varepsilon)p$. 
In the end, we take the expansion around $\varepsilon \approx 0$ and keep only the zeroth order term, in 
the absence of divergences. The result is finite and can be elegantly written in terms of the two-point 
function of the current operator,
\begin{align}
\llangle T_{\mu\nu}(0)J_{\rho}^{a}(\bar{p})J_{\sigma}^{b}(-\bar{p})\rrangle & = \delta_{\mu \nu} \llangle 
J^{a}_{\rho}(\bar{p}) J^{b}_{\sigma}(-\bar{p}) \rrangle - \delta_{\mu \sigma} \llangle J^{a}_{\rho}(\bar{p}) 
J^{b}_{\nu}(-\bar{p}) \rrangle \nonumber \\ 
& \qquad \quad- \delta_{\mu \rho} \llangle J^{a}_{\sigma}(\bar{p}) J^{b}_{\nu}(-\bar{p}) \rrangle  - \bar{p}
_{\nu} \frac{\partial}{\partial \bar{p}^{\mu}} \llangle J^{a}_{\rho}(\bar{p}) J^{b}_{\sigma}(-\bar{p}) \rrangle + 
\mathcal{O}(\epsilon) \,. \label{TJJ_squeezed_limit}
\end{align}
in accordance with the analysis on soft limits in holographic cosmology done in \cite{HCsoft}.

\subsection{Holographic predictions} 

We are finally ready to apply our result to the holographic formulas 
\eqref{holographic_formula_3pt_function_tensor}, and obtain the corresponding cosmological correlators 
Using the result \eqref{TJJ_squeezed_limit}, it is useful to write the 3-point function $\langle TJJ \rangle$ 
as 
\begin{align}
    & -\text{Re}[(-i)^{d}\llangle T_{\mu\nu}(0)J^{a}_{\rho}(\bar{p})J^{b}_{\sigma}(-\bar{p})\rrangle]  =    
    \nonumber \\ 
    &  \hspace{1.5cm} \delta_{\mu\nu} \left\{ -\text{Re}[(-i)^{d}\llangle J^{a}_{\rho}(\bar{p})J^{b}_{\sigma}(-
    \bar{p})\rrangle\right\}-\delta_{\mu\sigma} \left\{ -\text{Re}[(-i)^{d}\llangle J^{a}_{\rho}(\bar{p})J^{b}_{\nu}
    (-\bar{p})\rrangle\right\} \nonumber \\ 
    & \hspace{2.5cm}-\delta_{\mu\rho}\left\{ -\text{Re}[(-i)^{d}\llangle J^{a}_{\sigma}(\bar{p})J^{b}_{\nu}(-
    \bar{p})\rrangle\right\}  -p_{\nu}\frac{\partial}{\partial p^{\mu}}\left\{ -\text{Re}[(-i)^{d}\llangle J^{a}_{\rho}
    (\bar{p})J^{b}_{\sigma}(-\bar{p})\rrangle\right\}, 
\end{align}
still in terms of the electric current. To write the corresponding version of this expression in terms of the 
vortex current, we apply the $Sl(2,\mathbb{Z})$ duality on both sides, and using once again the 
holographic formula for the 2-point function \eqref{holo_formula_AA}, we write the following expression for 
$\langle T \tilde{J} \tilde{J} \rangle$ in the squeezed limit: 
\begin{align}
    & -2\text{Re}[(-i)^{d}\llangle T_{\mu\nu}(0) \tilde J^{a}_{\rho}(\bar{p}) \tilde J^{b}_{\sigma}(-\bar{p})
    \rrangle]=   \nonumber \\ 
    & \hspace{1cm} \frac{\delta^{ab}}{\mathcal{P}_{\tilde{A}}(p)}\bigg[\delta_{\mu\nu}\pi_{\rho\sigma}(p)-
    \delta_{\mu\sigma}\pi_{\rho\nu}(p)-\delta_{\mu\rho}\pi_{\sigma\nu}(p) - \mathcal{P}_{\tilde{A}}(p)p_{\nu}
    \frac{\partial}{\partial p^{\mu}}\left(\frac{\pi_{\rho\sigma}(p)}{\mathcal{P}_{\tilde{A}}(p)}\right)\bigg]\, . 
    \label{TJJ_squeezed_limit_analyt_continued}
\end{align}

Finally, take the squeezed limit of the formula \eqref{main_holo_formula_xi_AA} and replace the result 
\eqref{TJJ_squeezed_limit_analyt_continued}, to obtain  
\begin{align}\label{result_xi}
    & \llangle\xi(0)\tilde{A}^{a}_{\rho}(p)\tilde{A}^{b}_{\sigma}(-p)\rrangle  \nonumber \\
    & \hspace{2cm} = -2\mathrm{Re}\left[(-i)^{d}\delta^{\mu\nu}\llangle T_{\mu\nu}(0)\tilde{J}
    ^{a\lambda_{3}}(\bar{p})\tilde{J}^{b\lambda_{4}}(-\bar{p})\rrangle\right]  \pi_{\rho\lambda_{3}}(p)
    \pi_{\sigma\lambda_{4}}(p)\mathcal{P}_{\xi}(0)\mathcal{P}^2_{\tilde A}(p) \nonumber \\
    &\hspace{2cm} =\delta^{ab}\left(1+p\frac{\mathcal{P}^{\prime}_{\tilde A}(p)}{\mathcal{P}_{\tilde{A}}(p)}
    \right) 
    \pi_{\rho\sigma}(p)\mathcal{P}_{\xi}(0)\mathcal{P}_{\tilde{A}}(p) \, .
\end{align}
Note that this formula has precisely the expected form from cosmology 
\eqref{factorization_squeezed_limit}, with
\begin{equation}
    \mathcal{D}(p) = 1+p\frac{\mathcal{P}^{\prime}_{\tilde A}(p)}{\mathcal{P}_{\tilde{A}}(p)} = 1 + \frac{d \log 
    \mathcal{P}_{\tilde A}(p)}{d \log p}
\end{equation}
giving the ``running scaling" of the magnetic monopole power spectrum. We should keep in mind, 
however, that this is a schematic formula. The functional form of this formula, being proportional to the 
product of the power spectrum of $\xi$ and $\tilde A$, depends only on the fact that the three-point 
function $\langle T \tilde J \tilde J \rangle$ becomes proportional to $ \langle \tilde J  \rangle$ and 
its derivative in the squeezed limit. This behavior is expected to hold even at higher-loop orders. 

Replacing the explicit holographic results at 1-loop, we find 
\begin{align}\label{zero_xi}
    \lim_{p_1 \to 0} \llangle\xi(p_1)\tilde{A}^{a}_{\rho}(p)\tilde{A}^{b}_{\sigma}(-p)\rrangle & = \delta^{ab} 
    \left( 1 + p \frac{\mathcal{P}'_{\tilde A}(p)}{\mathcal{P}_{\tilde{A}}(p)} \right) 
    \pi_{\rho\sigma}(p) \mathcal{P}
    _{\xi}(p_1) \mathcal{P}_{\tilde{A}}(p) \nonumber \\
    & = \delta^{ab} \left( 1 - 1 \right) \pi_{\rho\sigma}(p) \left( \frac{32}{13 N^2 p_1^3} \right) \left( -\frac{\pi^2 
    N^2}{2p} \right) \nonumber \\
    & = 0 \, .
\end{align}
This means that at 1-loop, there is no breaking of diffeomorphism invariance induced by the insertion of 
scalar perturbations. If the structure \eqref{TJJ_squeezed_limit} remains true at higher loop orders as 
expected, quantum corrections to the two-point function of the current operator can induce a nonzero 
result for the correlator $\langle \xi \tilde A \tilde A \rangle$ in the squeezed limit. At 2-loops, the power 
spectrum  $\mathcal{P}_{\tilde{A}}(p)$ has the leading-order form 
\begin{equation}
    \mathcal{P}_{\tilde{A}}(p) = -\frac{\pi^{2}N^{2}}{2p}\left( 1+\frac{g^2_{YM}N}{p} c\right)
\end{equation}
then, assuming \eqref{TJJ_squeezed_limit} holds, the result would be
\begin{equation}
    \mathcal{D}(p) \simeq -\frac{g_{YM}^2N c}{p} \left( 1 - \frac{g_{YM}^2N c}{p} \right) \simeq 
    -\frac{g_{YM}^2N c}{p} + \mathcal{O}(g_{YM}^4).
\end{equation}

It remains to do the same for the $\langle \gamma \tilde A \tilde A \rangle$ correlator. We first take the 
squeezed limit of the formula \eqref{main_holo_formula_gamma_AA},
\begin{align}
    \llangle\gamma_{\mu\nu}(0)\tilde{A}^{a}_{\rho}(p)\tilde{A}^{a}_{\sigma}(-p)\rrangle & =-\frac{1}
    {2}\mathrm{\text{Re}}\left[(-i)^{d}\llangle T^{\alpha\beta}(0)\tilde{J}^{a\lambda_{3}}(\bar{p})\tilde{J}
    ^{d\lambda_{4}}(-\bar{p})\rrangle\right] \nonumber \\ 
    & \hspace{2cm} \times \pi_{\rho\lambda_{3}}(p)\pi_{\sigma\lambda_{4}}(p)\Pi_{\mu\nu\alpha\beta}
    (0)\mathcal{P}_{\gamma}(0)\mathcal{P}^{2}_{A}(p),
\end{align}
and then we use the result \eqref{TJJ_squeezed_limit_analyt_continued} to write, after all simplifications 
\begin{align}
    \llangle\gamma_{\mu\nu}(0)\tilde{A}^{a}_{\rho}(p)\tilde{A}^{b}_{\sigma}(-p)\rrangle & = \frac{\delta^{ab}}
    {4}\bigg( \delta^{\alpha\beta}\pi_{\rho\sigma}(p)-\pi_{\rho}{}^{\beta}(p)\pi_{\sigma}{}^{\alpha}(p)-\pi_{\rho}
    {}^{\alpha}(p)\pi_{\sigma}{}^{\beta}(p) \nonumber \\
    & \hspace{2cm}+ \frac{p^{\beta}}{\mathcal{P}_{\tilde{A}}(p)}\frac{\partial\mathcal{P}_{\tilde{A}}(p)}
    {\partial p_{\alpha}}\pi_{\rho\sigma}(p)\bigg) \Pi_{\mu\nu\alpha\beta}(0)\mathcal{P}_{\gamma}
    (0)\mathcal{P}_{\tilde{A}}(p).
\end{align}
This formula seems much more complicated then the one for the insertion of the scalar perturbation due 
to the tensor structure of the tensor perturbation. We can simplify the result to extract the physics by 
contracting $\mu$ with $\rho$ and $\nu$ with $\sigma$ (in analogy to the way we calculate the scalar 
tensor power spectrum), to find
\begin{align}\label{int_result_gamma_AA}
    \llangle\gamma_{\mu\nu}(0)\tilde{A}^{a\mu}(p)\tilde{A}^{b\nu}(-p)\rrangle & = \frac{\delta^{ab}}
    {4}\mathcal{P}_{\gamma}(0)\mathcal{P}_{\tilde{A}}(p)\Bigg[-(\pi_{\mu\nu}(0)\pi^{\mu\nu}(p))^{2} 
    \nonumber \\
    & \hspace{-2cm}+ \frac{\mathcal{P}_{\tilde{A}}'(p)}{p\mathcal{P}_{\tilde{A}}(p)}\left(p^{\alpha}
    \pi_{\alpha\mu}(0)\pi^{\mu\nu}(p)\pi_{\nu\beta}(0)p^{\beta}-\frac{1}{2}\pi_{\mu\nu}(0)\pi^{\mu\nu}
    (p)p^{\alpha}\pi_{\alpha\beta}(0)p^{\beta}\right)\Bigg]. 
\end{align}
To simplify this equation even further, we evaluate the transverse projector for the $p_1$ before taking the 
limit, using the parametrization \eqref{def_squeezed_limit} (note that here we will use the momentum 
variables for cosmology, after analytical continuation, since the transverse projector is invariant under $
\bar{p} \rightarrow i p$):
\begin{equation}
    \pi_{\mu\nu}(p_1) = \delta_{\mu\nu} - \frac{p_{1\mu}p_{1\nu}}{p_1^2} = \delta_{\mu\nu} - 
    \frac{(2\varepsilon p_\mu)(2\varepsilon p_\nu)}{4\varepsilon^2 p^2} = \delta_{\mu\nu} - \frac{p_\mu 
    p_\nu}{p^2} \, .
\end{equation}
Because the $\varepsilon$ dependence cancels exactly, the limit $\varepsilon \to 0$ is finite. Using this 
particular choice of collinear squeezed limit, we find
\begin{equation}
    \pi_{\mu\nu}(0) = \delta_{\mu\nu} - \frac{p_\mu p_\nu}{p^2} = \pi_{\mu\nu}(p) \, .
\end{equation}
With this result, the tensor contractions in the bispectrum simplify drastically. In $d=3$ spatial dimensions, 
the product of two transverse projectors yields:
\begin{equation}
    \pi_{\mu\nu}(p)\pi^{\mu\nu}(p) = d - 1 = 2 \, .
\end{equation}
Consequently,
\begin{equation}
    -(\pi_{\mu\nu}(0)\pi^{\mu\nu}(p))^{2} = -(\pi_{\mu\nu}(p)\pi^{\mu\nu}(p))^{2} = -(2)^2 = -4 \, .
\end{equation}

For the terms that multiply the derivative of the power spectrum $\mathcal{P}_A$, we use the fact that a 
transverse projector annihilates its corresponding momentum:
\begin{equation}
    p^\alpha \pi_{\alpha\mu}(0) = p^\alpha \pi_{\alpha\mu}(p) = p^\alpha \left( \delta_{\alpha\mu} - 
    \frac{p_\alpha p_\mu}{p^2} \right) = p_\mu - p_\mu = 0 \,,  
\end{equation}
which means that the whole dependence on the dimensional running of the bispectrum is zero. The full 
expression collapses to:
\begin{equation}
    \llangle\gamma_{\mu\nu}(0)\tilde{A}^{a\mu}(p)\tilde{A}^{b\nu}(-p)\rrangle = -\delta^{ab}\mathcal{P}
    _{\gamma}(0)\mathcal{P}_{\tilde{A}}(p) \, .
\end{equation}
where the hard divergence when $p_1 \rightarrow 0$ remains present on the power spectrum $
\mathcal{P}_{\gamma}(0)$.

However, as we said before, the parametrization for the squeezed limit \eqref{def_squeezed_limit} forces 
the $p_{1\mu} \rightarrow 0$ to be taken from a particular direction in space, namely, all three momenta 
are collinear from the start. In a cosmological setup, the long mode $p_1 \to 0$ can originate from any 
angle relative to the hard momentum $p$. For completeness, we will simplify the equation 
\eqref{int_result_gamma_AA} in a different way to capture this, by leaving $p_1$ explicit and defining the 
unit vector $\hat{p}_1^\mu = p_1^\mu / p_1$. Then:
\begin{equation}
    \pi_{\mu\nu}(0) = \delta_{\mu\nu} - \hat{p}_{1\mu} \hat{p}_{1\nu} \, .
\end{equation}
Let $\theta$ be the angle between $\hat{p}_1$ and the hard momentum $p$. The inner product is 
therefore $\hat{p}_1 \cdot p = p \cos\theta$, which means that now the direction from which we collapse 
the shape of the triangle formed by $p_{1\mu} = p_{2\mu} + p_{3\mu}$ is completely generic. One may 
wonder if this is really generic, since we already use the parametrization \eqref{def_squeezed_limit} in 
\eqref{TJJ_squeezed_limit}. The answer is yes, since the result \eqref{TJJ_squeezed_limit} is finite and 
independent of $\varepsilon$ (or $p_1$), meaning we would obtain the same result by just taking the limit 
$p_1 \rightarrow 0$ without any particular parametrization. The only new dependence on $p_1$ that 
appears in \eqref{int_result_gamma_AA} came from the $\langle T T \rangle(p_1)$ propagator in the 
denominator of the holographic formulas. 

With this choice, we find (in 3 dimensions)
\begin{align}
    \pi_{\mu\nu}(0)\pi^{\mu\nu}(p) &= \left( \delta_{\mu\nu} - \hat{p}_{1\mu} \hat{p}_{1\nu} \right) 
    \left( \delta^{\mu\nu} - \frac{p^\mu p^\nu}{p^2} \right) \nonumber \\
    &=  1 + \cos^2\theta \, .
\end{align}
and
\begin{equation}
    p^\alpha \pi_{\alpha\beta}(0) p^\beta = p^2 - (\hat{p}_1 \cdot p)^2 = p^2(1 - \cos^2\theta) = p^2 
    \sin^2\theta \, .
\end{equation}
Next, letting $v_\mu = p^\alpha \pi_{\alpha\mu}(0)$, we contract $v_\mu \pi^{\mu\nu}(p) v_\nu$:
\begin{equation}
    v_\mu \pi^{\mu\nu}(p) v_\nu = (v \cdot v) - \frac{(v \cdot p)^2}{p^2} = p^2 \cos^2\theta \sin^2\theta \, .
\end{equation}
The term inside parentheses in eq. (\ref{int_result_gamma_AA}) then becomes
\begin{align}
    \left( v_\mu \pi^{\mu\nu}(p) v_\nu - \frac{1}{2} \pi_{\mu\nu}(0)\pi^{\mu\nu}(p) p^\alpha \pi_{\alpha\beta}(0) 
    p^\beta \right) &= p^2 \cos^2\theta \sin^2\theta - \frac{1}{2}(1 + \cos^2\theta)p^2 \sin^2\theta \nonumber 
    \\
    &= -\frac{1}{2} p^2 \sin^4\theta \, .
\end{align}
Substituting these results back into the full correlator, the $p^2$ cancels the factor of $p$ in the 
denominator, and we find an angle-dependent result:
\begin{equation}
    \llangle\gamma_{\mu\nu}(0)\tilde{A}^{a\mu}(p)\tilde{A}^{b\nu}(-p)\rrangle = \frac{\delta^{ab}}
    {4}\mathcal{P}_{\gamma}(0)\mathcal{P}_{\tilde{A}}(p) \left[ -(1 + \cos^2\theta)^2 - \frac{1}{2} p 
    \frac{\mathcal{P}_{\tilde{A}}'(p)}{\mathcal{P}_{\tilde{A}}(p)} \sin^4\theta \right] \, .
\end{equation}

To extract the physical result observed in cosmology, we can take the average over all possible 
orientations of the long-wave mode $\hat{p}_1$ in three-dimensional momentum space. Using the solid 
angle averages $\langle \cos^2\theta \rangle = 1/3$ and $\langle \cos^4\theta \rangle = 1/5$, we find:
\begin{equation}
    \langle (1 + \cos^2\theta)^2 \rangle = \frac{28}{15}, \qquad \langle \sin^4\theta \rangle = \langle (1 - 
    \cos^2\theta)^2 \rangle = \frac{8}{15} \, .
\end{equation}
Applying this to our result, we finally obtain the most general squeezed limit
\begin{equation}\label{result_gamma}
    \llangle\gamma_{\mu\nu}(0)\tilde{A}^{a\mu}(p)\tilde{A}^{b\nu}(-p)\rrangle_{\theta-\text{average}} = -
    \frac{\delta^{ab}}{15}\mathcal{P}_{\gamma}(0)\mathcal{P}_{\tilde{A}}(p) \left( 7 + p \frac{\mathcal{P}
    _{\tilde{A}}'(p)}{\mathcal{P}_{\tilde{A}}(p)} \right) \, ,
\end{equation}
and replacing the holographic results for $P_{\tilde A}(p), -\pi^2N^2/(2p)$, as in \eqref{zero_xi}, we find
\begin{equation}\label{result_gamma1}
    \llangle\gamma_{\mu\nu}(0)\tilde{A}^{a\mu}(p)\tilde{A}^{b\nu}(-p)\rrangle_{\theta-\text{average}} = -
    \frac{6}{15} \mathcal{P}_{\gamma}(0)\mathcal{P}_{\tilde{A}}(p) \delta^{ab} \, ,
\end{equation}

\section{Conclusion and Discussion}\label{sec:discussion}

In this paper, we have computed the non-Gaussianities associated with the cross-correlation between 
monopole and metric perturbations by applying the holographic dictionary proposed in \cite{holouniverse} 
to the $\langle TJJ \rangle $ correlator of a 3d QFT. This work follows the path of 
\cite{holonongauss,cosm3pt,holopredic,matheuspaper} in which holographic non-Gaussianities were 
computed for the metric perturbations distribution and the monopole field distribution. We can perform a 
similar analysis to our results as is done for the bispectra in those cases, and in cosmology in general. 
Let us take the result for the $\langle \xi A A\rangle$ correlator. Following the schematic equation 
\eqref{def_shapefunction}, we can write
\begin{equation}
   \lim_{p_1\rightarrow 0} \llangle \xi(p_1) \tilde A(p) \tilde A (-p) \rrangle \sim f_{NL}^{\xi \tilde A \tilde A}(p)  
   \lim_{p_1\rightarrow 0} \mathcal{P}_{\xi}(p_1) \mathcal{P}_{\tilde A} (p),
\end{equation}
where, using the equality \eqref{equality_D_f_NL}, we have
\begin{equation}
   f_{NL}^{\xi \tilde{A} \tilde{A}}= 1+p\frac{\mathcal{P}^{\prime}_{\tilde{A}}(p)}{\mathcal{P}_{\tilde{A}}(p)}\, ,
\end{equation}
as the effective non-linear parameter for the scalar-monopole cross-correlation in the squeezed limit. As 
noted, this term captures any non-linear deviations from Gaussianity in the monopole field induced by the 
background scalar perturbation $\xi$. Our holographic results yield $f_{NL}^{\xi \tilde{A} \tilde{A}}=0$, 
implying that no such modulation occurs at leading order.

We can do the same for the result obtained for $\langle \gamma \tilde{A} \tilde{A}\rangle$ in the squeezed 
limit \eqref{result_gamma}, writing
\begin{equation}
    \lim_{p_1\rightarrow 0} \llangle \gamma(p_1) \tilde A(p) \tilde A (-p) \rrangle \sim f_{NL}^{\gamma \tilde A 
    \tilde A}(p)  \lim_{p_1\rightarrow 0} \mathcal{P}_{\gamma}(p_1) \mathcal{P}_{\tilde A} (p),
\end{equation}
from which we find the tensor-monopole cross-correlation parameter, after taking the average over the 
angles of the soft momentum $p_1$, to be
\begin{equation}
    f_{NL}^{\gamma \tilde{A} \tilde{A}} = 7 + p \frac{\mathcal{P}_{\tilde{A}}'(p)}{\mathcal{P}_{\tilde{A}}(p)},
\end{equation}
carrying the contribution for the modulation of the cosmological correlator analogous to the scalar case, 
but now induced by the tensor metric perturbation $\gamma_{\mu\nu}$. From our results, $f_{NL}
^{\text{\small$\gamma \tilde{A} \tilde{A}$}} =6$. 

The factorization of the correlator into a shape function times the power spectra, allowed by taking the 
squeezed limit, also permits us to evaluate the amplitude hierarchy of the non-Gaussian correlators. 
Comparing our results \eqref{result_gamma} and \eqref{result_xi} to the general factorization 
\eqref{factorization_squeezed_limit}, we conclude that
\begin{equation}
    \langle \xi \tilde{A} \tilde{A} \rangle  \,\,\gg \,\, \langle \gamma \tilde{A} 
    \tilde{A} \rangle\;,
\end{equation}
since observational bounds impose the constraint 
$\mathcal{P}_{\gamma} \ll \mathcal{P}_{\xi}$. Note that, while the dimensionless power spectra 
naively satisfy $\tilde P_\xi\ll \tilde P_{\tilde A}$, this must be due to the fact that $\tilde P_{\tilde A}$
measures mainly the size of the monopoles relative to the cosmological scale $p$, and only the 
non-Gaussianity factor $f_{NL}$ is relevant, in the absence of observations of monopoles.

Note that the results we found in the squeezed limit are actually divergent as $p_1 \rightarrow 0$, since 
both $\mathcal{P}_\xi$ and $\mathcal{P}_{\tilde{A}}$ scale as $p_1^{-3}$. To get a finite result, we should 
take $p_1$ to be small but non-zero, or follow the calculations of \cite{matheusIRdivergences} and 
consider a mass-deformed toy model for cosmological calculations.

Finally, we would like to comment on the addition of the contact terms \eqref{CTP} to obtain the full 
correlator in the 3-dimensional QFT. Even though the procedure of computing correlation functions in QFT 
is standardized by calculating the connected diagrams using Feynman rules, the different conventions for 
the definition of $n$-point functions in \eqref{TJJfulldef} and \eqref{TJJflatdef} may confuse. In 
holographic cosmology, the definition \eqref{TJJfulldef} must be the one to be applied to find the dual 
cosmological operators for two reasons. First, the dictionary is first derived for one-point functions of 
boundary operators, and the map for higher $n$-point functions is obtained by taking further functional 
derivatives with respect to the sources. This coincides with the definition \eqref{TJJfulldef}. Second, 
without the contact terms in \eqref{CTP}, we would not be able to find the expected factorization of the 
cosmological non-Gaussianities in the squeezed limit. That is because these terms are semi-local, i.e., 
they account for the insertion of operators/bulk fields in coincident points in space, precisely contributing 
to the behaviour of the correlation function when one of the modes is much smaller than the others.

Proceeding from these considerations, an interesting direction to pursue is to find sub-leading 
contributions to $\langle \xi \tilde{A} \tilde{A} \rangle$ to check if the result $f_{NL}^{\xi \tilde{A} \tilde{A}}
=0$ holds at higher orders. We showed that, if the factorization \eqref{TJJ_squeezed_limit} remains the 
same at 2-loop, we have a non-zero contribution proportional to the effective dimensionless coupling, that 
is, 
\begin{equation}
    f_{NL}^{\xi \tilde{A} \tilde{A}}(p) = \mathcal{O}(g^2_{\text{eff}}), \qquad g^2_{\text{eff}} = \frac{g^2_{YM}
     N}{p}.
\end{equation}
Exploring if this is true and calculating these contributions is left for future work.  

Another interesting direction would be to revisit $\langle T T J \rangle$, which, according to the general 
factorization \eqref{factorization_squeezed_limit}, is expected to give a greater amplitude than $\langle 
TJJ \rangle$, and yet, it is naively zero due to the global symmetry of our toy model. To avoid that, one 
could consider a model with a $U(1)$ conserved current rather than the $SO(3)$ considered here. While 
this does not directly map the results to monopole configurations in cosmology, the $U(1)$ global 
symmetry alone may be enough to capture the dynamics and give an approximation for $f_{NL}^{\xi \xi 
\tilde A }$,  which ideally should be calculated using non-abelian vortex solutions dual to 't Hooft 
monopoles in cosmology.

\section*{Acknowledgments}

The work of  JZF is supported by FAPESP grant 2025/06199-0. MC is supported by FAPESP grant
 2022/02791-4.
The work of HN is supported in part by  CNPq grant 
304583/2023-5 and FAPESP grant  2024/15298-0.
HN would also like to thank the ICTP-SAIFR for their support 
through FAPESP grant 2021/14335-0.

\appendix 

\section{Contact terms and Ward Identity}\label{appendA}

Here we derive the contact terms presented in \eqref{CTP} and the transverse Ward identity that the QFT 
correlator, when summed with such terms, should satisfy. 

\subsection{Contact terms to the $\left\langle TJJ\right\rangle$ correlator}

To obtain the three-point function $\left\langle TJJ\right\rangle$, we may start with the one-point function 
for the stress-energy tensor
\begin{equation}
    \langle T_{\mu\nu}(x_{1})\rangle=-\frac{2}{\sqrt{g(x_{1}})}\frac{\delta}{\delta g^{\mu\nu}}Z[g,a],
\end{equation}
and take functional derivatives with respect to the source of the current $J_\mu$, 
\begin{equation}\label{3pt}
    \left\langle T_{\mu\nu}(x_{1})J_{\rho}(x_{2})J_{\sigma}(x_{3})\right\rangle  = \left(\frac{-1}{\sqrt{g(x_{3})}}
    \frac{\delta}{\delta a^{\sigma}(x_{3})}\right)\left(\frac{-1}{\sqrt{g(x_{2})}}\frac{\delta}{\delta a^{\rho}(x_{2})}
    \right) \langle T_{\mu\nu}(x_{1})\rangle.
\end{equation}
This is equivalent to \eqref{TJJfulldef}. By considering 
\begin{equation}
    Z\left[g_{\mu \nu} , a_\mu \right]= \int\mathcal{D}\Phi e^{-S - \int d^{d}x\sqrt{g}~a\cdot J},
\end{equation}
where $S$ is the toy model action \eqref{toymodel}, and applying the functional derivatives in \eqref{3pt}, 
we find
\begin{align}
& \left\langle T_{\mu\nu}(x_{1})J_{\rho}(x_{2})J_{\sigma}(x_{3})\right\rangle = \nonumber\\
    &=\int\mathcal{D}\Phi e^{-S_{\text{tot}}}J_{\rho}(x_{2})J_{\sigma}(x_{3})\Big[ \frac{2}{\sqrt{g(x_{1})}}
    \frac{\delta}{\delta g^{\mu\nu}(x_{1})}S+a_{\mu}(x_{1})J_{\nu}(x_{1})+a_{\nu}(x_{1})J_{\mu}(x_{1}) . 
    \nonumber\\
    & \hspace{9.455cm} -g_{\mu\nu}(x_{1})g^{\tau\xi}(x_{1})a_{\tau}(x_{1})J_{\xi}(x_{1}) \Big] \nonumber\\
    & +\int\mathcal{D}\Phi e^{-S_{\text{tot}}}J_{\rho}(x_{2})\Big[ \frac{-1}{\sqrt{g(x_{3})}}\frac{\delta}{\delta 
    a^{\sigma}(x_{3})}\frac{2}{\sqrt{g(x_{1})}}\frac{\delta}{\delta g^{\mu\nu}(x_{1})}S-\frac{1}{\sqrt{g(x_{3})}}
    g_{\mu\sigma}\delta(x_{1}-x_{3})J_{\nu}(x_{1})  \nonumber\\
    & \hspace{3.65cm} -\frac{1}{\sqrt{g(x_{3})}}g_{\nu\sigma}\delta(x_{1}-x_{3})J_{\mu}(x_{1})+\frac{1}
    {\sqrt{g(x_{3})}}\delta(x_{1}-x_{3})g_{\mu\nu}(x_{1})J_{\sigma}(x_{1}) \Big] \nonumber\\
    & +\int\mathcal{D}\Phi e^{-S_{\text{tot}}}J_{\sigma}(x_{3})\Big[\frac{-1}{\sqrt{g(x_{2})}}\frac{\delta}{\delta 
    a^{\rho}(x_{2})}\frac{2}{\sqrt{g(x_{1})}}\frac{\delta}{\delta g^{\mu\nu}(x_{1})}S-\frac{1}{\sqrt{g(x_{2})}}
    g_{\mu\rho}\delta(x_{1}-x_{2})J_{\nu}(x_{1})  \nonumber\\
    & \hspace{3.65cm} -\frac{1}{\sqrt{g(x_{2})}}g_{\nu\rho}\delta(x_{2}-x_{1})J_{\mu}(x_{1})+\frac{1}
    {\sqrt{g(x_{2})}}g_{\mu\nu}(x_{1})\delta(x_{1}-x_{2})J_{\rho}(x_{1})\Big] \nonumber\\
    & \hspace{2.3cm} +\int\mathcal{D}\Phi e^{-S_{\text{tot}}}\Big[\frac{-1}{\sqrt{g(x_{3})}}\frac{\delta}{\delta 
    a^{\sigma}(x_{3})}\frac{-1}{\sqrt{g(x_{2})}}\frac{\delta}{\delta a^{\rho}(x_{2})}\frac{2}{\sqrt{g(x_{1})}}
    \frac{\delta}{\delta g^{\mu\nu}(x_{1})}S \Big] .
\end{align}
Now we can take $a\rightarrow0 \text{ and } g\rightarrow0$ in the expression above 
\begin{align}
    & =\int\mathcal{D}\Phi e^{-S_{\text{tot}}}J_{\rho}(x_{2})J_{\sigma}(x_{3})T_{\mu\nu}(x_{1}) +
    \int\mathcal{D}\Phi e^{-S_{\text{tot}}}\left[\frac{\delta}{\delta a^{\sigma}(x_{3})}\frac{\delta}{\delta a^{\rho}
    (x_{2})}T_{\mu\nu}(x_{1})\right] \nonumber\\
    & +\int\mathcal{D}\Phi e^{-S_{\text{tot}}}J_{\rho}(x_{2})\Big[{-\frac{\delta}{\delta a^{\sigma}(x_{3})}
    T_{\mu\nu}(x_{1})}-\delta_{\mu\sigma}\delta(x_{1}-x_{3})J_{\nu}(x_{1})-\delta_{\nu\sigma}\delta(x_{1}-
    x_{3})J_{\mu}(x_{1}) \nonumber \\
    & \hspace{9.45cm} +\delta(x_{1}-x_{3})\delta_{\mu\nu}(x_{1})J_{\sigma}(x_{1})\Big] \nonumber\\
    &+\int\mathcal{D}\Phi e^{-S_{\text{tot}}}J_{\sigma}(x_{3})\Big[{-\frac{\delta}{\delta a^{\rho}(x_{2})}
    T_{\mu\nu}(x_{1})}-\delta_{\mu\rho}\delta(x_{1}-x_{2})J_{\nu}(x_{1})-\delta_{\nu\rho}\delta(x_{2}-
    x_{1})J_{\mu}(x_{1}) \nonumber \\
    & \hspace{8.9cm}+\delta_{\mu\nu}(x_{1})\delta(x_{1}-x_{2})J_{\rho}(x_{1})\Big],
\end{align}
where we applied the definition of the stress-energy tensor
\begin{equation}
    T_{\mu\nu}=\left.\frac{2}{\sqrt{g(x_{1}})}\frac{\delta}{\delta g^{\mu\nu}}S\right|_{g_{\mu\nu}
    =\delta_{\mu\nu}}.
\end{equation}

Finally, we can write the full three-point function as 
\begin{align}\label{ctx}
    & \left\langle T_{\mu\nu}(x_{1})J_{\rho}(x_{2})J_{\sigma}(x_{3})\right\rangle _{\text{full}} = {\left\langle J_{\rho}
    (x_{2})J_{\sigma}(x_{3})T_{\mu\nu}(x_{1})\right\rangle _{flat}}\nonumber\\
    & \hspace{2.5cm} -\delta_{\mu\sigma}\left\langle \delta(x_{1}-x_{3})J_{\rho}(x_{2})J_{\nu}(x_{1})
    \right\rangle -\delta_{\nu\sigma}\left\langle \delta(x_{1}-x_{3})J_{\rho}(x_{2})J_{\mu}(x_{1})\right\rangle 
    \nonumber  \\
    & \hspace{2.5cm}-\delta_{\mu\rho}\left\langle \delta(x_{1}-x_{2})J_{\sigma}(x_{3})J_{\nu}(x_{1})
    \right\rangle -\delta_{\nu\rho}\left\langle \delta(x_{2}-x_{1})J_{\sigma}(x_{3})J_{\mu}(x_{1})\right\rangle 
    \nonumber \\
    & \hspace{2.5cm}+\delta_{\mu\nu}\left\langle \delta(x_{1}-x_{3})J_{\rho}(x_{2})J_{\sigma}(x_{1})
    \right\rangle +\delta_{\mu\nu}\left\langle \delta(x_{1}-x_{2})J_{\sigma}(x_{3})J_{\rho}(x_{1})\right\rangle 
    \nonumber \\
    &  \hspace{2.65cm}- \left\langle\frac{\delta^2 T_{\mu\nu}(x_1) }{\delta a_\sigma(x_3)\delta a_\rho(x_2)}
    \right\rangle - \left\langle J_{\rho}(x_{2})\frac{\delta T_{\mu\nu}(x_{1})}{\delta a^{\sigma}(x_{3})}
    \right\rangle 
    -\left\langle J_{\sigma}(x_{3})\frac{\delta T_{\mu\nu}(x_{1})}{\delta a^{\rho},(x_{2})}\right\rangle \, , 
\end{align}
where 
\begin{equation}
  \left\langle J_{\rho}(x_{2})J_{\sigma}(x_{3})T_{\mu\nu}(x_{1})\right\rangle_{\text{flat}} = \int\mathcal{D}
  \Phi e^{-S_{\text{tot}}}J_{\rho}(x_{2})J_{\sigma}(x_{3})T_{\mu\nu}(x_{1}).
\end{equation}

The terms in the second, third, and fourth lines are the contact terms that we must add in order to get the 
full correlator. In our case, the last two terms are zero because our gauge field works just as a source for 
the current $J$; it is not a degree of freedom. The first one on the last line is a \textit{ultralocal} term, 
meaning that it corresponds to the spatial coincidence of the three operators. We may drop it because 
such terms are scheme-dependent, meaning that they may be removed by the addition of local 
counterterms.

Since our calculations are in momentum space, for \eqref{ctx} we get 
\begin{align}\label{ctp}
    \llangle T_{\mu\nu}(p_{1})J_{\rho}(p_{2})J_{\sigma}(p_{3})\rrangle_{\text{full}} 
    &=\llangle T_{\mu\nu}(p_{1})J_{\rho}(p_{2})J_{\sigma}(p_{3})\rrangle_{\text{flat}} \notag\\
    &-2\delta_{\rho(\mu}\llangle J_{\nu)}(-p_{3})J_{\sigma}(p_{3})\rrangle-2\delta_{\sigma(\mu}\llangle 
    J_{\nu)}(-p_{2})J_{\rho}(p_{2})\rrangle \notag\\
    &+\delta_{\mu\nu}\left(\llangle J_{\rho}(-p_{3})J_{\sigma}(p_{3})\rrangle+\llangle J_{\sigma}(-
    p_{2})J_{\rho}(p_{2})\rrangle \right).
\end{align}

\subsection{Transverse Ward identity}

Ward identities for three-point functions in momentum space were derived within the framework of 
conformal field theories in \cite{3ptward}. Although our current setup is not conformally invariant, the 
energy-momentum tensor remains a conserved current, ensuring that the transverse Ward identity retains 
the same form. Below, we review this identity for the $\langle TJJ \rangle$ correlation function.

To derive the Ward identity associated to the conservation of the
energy-momentum tensor, we must use the fact that, for a theory coupled
to a dynamical metric $g$, the energy momentum tensor is associated
with the invariance of the theory under diffeomorphisms:
\begin{align*}
\delta_{\epsilon}g^{\mu\nu} & =-\nabla^{\mu}\epsilon^{\nu}-\nabla^{\nu}\epsilon^{\mu},\\
\delta_{\epsilon}a^{\mu} & =\epsilon^{\lambda}\nabla_{\lambda}a^{\mu}+a^{\lambda}\nabla^{\mu}
\epsilon_{\lambda}\, .
\end{align*}

The invariance of the generating functional $\delta_{\epsilon}Z[g,a]$
implies 

\begin{equation}
\delta_{\epsilon}Z[g,a]=\int d^{d}y\left[\left(\frac{\delta}{\delta g^{\mu\nu}(y)}Z[g,a]\right)\delta_{\epsilon}
g^{\mu\nu}(y)+\left(\frac{\delta}{\delta a^{\mu}(y)}Z[g,a]\right)\delta_{\epsilon}a^{\mu}(y)\right]\, .
\end{equation}
Applying the definitions 
\begin{align*}
\langle T_{\mu\nu}(x)\rangle & =-\frac{2}{\sqrt{g(x})}\frac{\delta}{\delta g^{\mu\nu}(x)}Z[g,a],\\
\langle J_{\mu}(x)\rangle & =-\frac{1}{\sqrt{g(x)}}\frac{\delta}{\delta a^{\mu}(x)}Z[g,a],
\end{align*}
we have 
\begin{align*}
\delta_{\epsilon}Z[g,a] & =\int d^{d}y\sqrt{g(y)}\left[\frac{1}{2}\langle T_{\mu\nu}(y)\rangle \left(\nabla^{\mu}
\epsilon^{\nu}+\nabla^{\nu}\epsilon^{\mu}\right)(y)-\langle J_{\mu}(y)\rangle\left(\epsilon^{\lambda}
\nabla_{\lambda}a^{\mu}+a^{\lambda}\nabla^{\mu}\epsilon_{\lambda}\right)(y)\right]\, .
\end{align*}
Now, since the stress-energy tensor is symmetric, we can write 
\begin{equation}
\langle T_{\mu\nu}(y)\rangle\left(\nabla^{\mu}\epsilon^{\nu}+\nabla^{\nu}\epsilon^{\mu}\right)(y)=2\langle 
T_{\mu\nu}(y)\rangle \nabla^{\mu}\epsilon^{\nu},
\end{equation}
such that 
\begin{equation}
\delta_{\epsilon}Z[g,a]=\int d^{d}y\sqrt{g(y)}\left[\langle T_{\mu\nu}(y)\rangle \nabla^{\mu}\epsilon^{\nu}-
\langle J_{\mu}(y)\rangle\left(\epsilon^{\lambda}\nabla_{\lambda}a^{\mu}+a^{\lambda}\nabla^{\mu}
\epsilon_{\lambda}\right)(y)\right].
\end{equation}

There are two terms with derivatives acting on $\epsilon$. Integrating by parts, we get 
\begin{equation}
 =\int d^{d}y\sqrt{g(y)}\epsilon^{\nu}\left[-\nabla^{\mu}\langle T_{\mu\nu}(y)\rangle +\left(\nabla^{\mu}
 a_{\nu}(y)-\nabla_{\nu}a^{\mu}(y)\right)\langle J_{\mu}(y)\rangle+a_{\nu}(y)\nabla^{\mu}\langle J_{\mu}
 (y)\rangle\right].
\end{equation}

Note that the term $\nabla^{\mu}a_{\nu}(y)-\nabla_{\nu}a^{\mu}(y)$ is what one identifies with the field strength when
$J$ is associated with a gauge symmetry. This is not our case. Now,
imposing the invariance of the partition function
\begin{align}
\delta_{\epsilon}Z[g,a] =0 
\, \quad \Rightarrow \, \quad  \nabla^{\mu}\langle T_{\mu\nu}\rangle -\left(\nabla^{\mu}a_{\nu}-\nabla_{\nu}
a^{\mu}\right)\langle J_{\mu}\rangle-a_{\nu}\nabla^{\mu}\langle J_{\mu}\rangle=0
\end{align}

This is the Ward identity for our case. The diffeo-invariant covariant
derivatives $\nabla^{\mu}$ can be taken to be flat $\partial^{\mu}$(from now on we are just differentiating 
with respect to $a^{\mu}$ to
obtain the three-point function, so there is no more metric dependence). The last term is zero since $J$ is 
conserved, and we are left with
\begin{equation}
\partial^{\mu}\langle T_{\mu\nu}\rangle=\left(\partial^{\mu}a_{\nu}-\partial_{\nu}a^{\mu}\right)\langle 
J_{\mu}\rangle.\label{WardTJ}
\end{equation}
Now, to obtain the identity for the three-point function, we differentiate both
sides. For the left-hand side, we get
\begin{align*}
\partial_{1}^{\mu}\langle T_{\mu\nu}(x_{1})J_{\rho}(x_{2})J_{\sigma}(x_{3})\rangle_{\text{full}} & 
=\left(\frac{-1}{\sqrt{g(x_{3})}}\frac{\delta}{\delta a^{\sigma}(x_{3})}\right)\left(\frac{-1}{\sqrt{g(x_{2})}}
\frac{\delta}{\delta a^{\rho}(x_{2})}\right)\partial_{1}^{\mu}\langle T_{\mu\nu}(x_{1})\rangle.
\end{align*}
Again, since we already differentiated with respect to the metric, we can set
it to flat, so (\ref{WardTJ}) becomes
\begin{align}\label{ward}
\partial_{1}^{\mu}\langle T_{\mu\nu}(x_{1})J_{\rho}(x_{2})J_{\sigma}(x_{3})\rangle_{\text{full}} & 
=\frac{\delta}{\delta a^{\sigma}(x_{3})}\frac{\delta}{\delta a^{\rho}(x_{2})}\partial_{1}^{\mu}\langle 
T_{\mu\nu}(x_{1})\rangle \nonumber \\
 & =\frac{\delta}{\delta a^{\sigma}(x_{3})}\frac{\delta}{\delta a^{\rho}(x_{2})}\left[\left(\partial_{1}^{\mu}
 a_{\nu}(x_{1})-\partial_{\nu}^{1}a^{\mu}(x_{1})\right)\langle J_{\mu}(x_{1})\rangle\right].
\end{align}

The left-hand side is what we derived in the last section, and results in \eqref{ctx}. Applying the functional 
derivatives to the right-hand side of \eqref{ward}, and then setting $a_\mu$ to zero, 
we obtain the transverse Ward identity for 
the full correlator,
\begin{align}\label{xward}
    \partial_{1}^{\mu}\langle T_{\mu\nu}(x_{1})J_{\rho}(x_{2})J_{\sigma}(x_{3})\rangle_{\text{full}} & =-
    \left\langle J_{\mu}(x_{1})J_{\sigma}(x_{3})\right\rangle \left(\partial_{1}^{\mu}\delta_{\nu\rho}
    \delta(x_{1}-x_{2})-\partial_{\nu}^{1}\delta_{\rho}^{\mu}\delta(x_{1}-x_{2})\right) \notag \\
    & \hspace{-0.5cm}-\left\langle J_{\mu}(x_{1})J_{\rho}(x_{2})\right\rangle \left(\partial_{1}^{\mu}
    \delta_{\sigma\nu}\delta(x_{1}-x_{3})-\partial_{\nu}^{1}\delta_{\sigma}^{\mu}\delta(x_{1}-x_{3})\right) \, .
\end{align}

In momentum space, this is equivalent to
\begin{align}
    &p_{1}^{\mu}\llangle T_{\mu\nu}(p_{1})J_{\rho}(p_{2})J_{\sigma}(p_{3})\rrangle_{\text{full}} \nonumber \\ 
    & \hspace{2cm} = 
    \left[-
    {2\delta_{\rho[\nu}p_{2}^{\mu]}\llangle J_{\mu}(-p_{3})J_{\sigma}(p_{3})\rrangle 
    -2\delta_{\sigma[\nu}p_{3}^{\mu]}\llangle J_{\mu}(-p_{2})J_{\rho}(p_{2})\rrangle }\right],
\end{align}
where we wrote the final expression in terms of the delta-stripped 3-point function as well, such that the $\delta^3(p_1-p_2-p_3)$ goes away \footnote{ This identity differs by an overall sign with the one displayed in \cite{3ptward}. This is due to the different momentum conventions: the referred paper works with all-incoming momenta $\sum_i p_i=0$, while we use $\bar{p}_1 = \bar{p}_2 + \bar{p}_3$. }. 

\subsection{Transverse projection}\label{transverse_proj}

The transverse projector associated to a momentum $p$ is defined as
\begin{equation}
	\pi_{\mu\nu}(p) = \delta_{\mu\nu} - \frac{p_\mu p_\nu}{p^2}, \qquad \Rightarrow \qquad \pi_{\mu\nu}(p) p^\nu = 0   .
	\label{eq:pidef}
\end{equation}
 
To obtain the transverse part of the full correlator, one must apply projectors with respect to the momenta involved in the three-point function, that is 
 \begin{equation}
    \llangle TJJ\rrangle_{\rm transv.} = \Pi^{\rm sym}_{\alpha\beta,\mu\nu}(\pb1) \pi^\rho_\gamma(\pb2) \pi^\sigma_\delta(\pb3)\times \llangle TJJ\rrangle_{\rm full} \notag    
 \end{equation}
 where 
 \begin{equation}
	\Pi^{\rm sym}_{\mu\nu,\alpha\beta}(p) \equiv \pi^\alpha_{(\mu}(p)\pi^\beta_{\nu)}(p) = \frac12\big(\pi_{\mu\alpha}(p)\pi_{\nu\beta}(p)+\pi_{\mu\beta}(p)\pi_{\nu\alpha}(p)\big) 
	\label{eq:Pisym}
\end{equation}
 ensures the projected object is symmetric under $(\mu,\alpha) \leftrightarrow (\nu,\beta)$. Note that this symmetrization has a trivial effect on the correlator, which is already symmetric under $\mu \leftrightarrow \nu$ by construction.

 Applying the projection over our result \eqref{fullcorrelator}, we find that exactly $15$ of the $28$ structures are annihilated identically, while each of the remaining $13$ structures is projected to an integer multiple of one of seven transverse building-block tensors, which we define as
\begin{equation}
	u_\alpha \equiv \pi_\alpha{}^\mu(\pb1) \pb{2\mu} ,\qquad v_\gamma \equiv \pi_\gamma{}^\rho(\pb2) \pb{3\rho} ,\qquad w_\delta \equiv \pi_\delta{}^\sigma(\pb3) \pb{2\sigma} ,
	\label{eq:uvw}
\end{equation}
and the transverse matrices
\begin{equation}
	P_{\alpha\gamma}\equiv\pi_\alpha{}^\mu(\pb1)\pi_{\mu\gamma}(\pb2) ,\qquad Q_{\alpha\delta}\equiv\pi_\alpha{}^\mu(\pb1)\pi_{\mu\delta}(\pb3) ,\qquad S_{\gamma\delta}\equiv\pi_\gamma{}^\rho(\pb2)\pi_{\rho\delta}(\pb3) .
	\label{eq:PQS}
\end{equation}
In terms of these, the seven tensor structures our transverse result can be written with respect to are
\begin{align}
	[R]_{1,\alpha\beta\gamma\delta} & \equiv u_\alpha u_\beta v_\gamma w_\delta ,                                         & [R]_{2,\alpha\beta\gamma\delta} & \equiv u_\alpha u_\beta S_{\gamma\delta} , \notag                                                 \\
	[R]_{3,\alpha\beta\gamma\delta} & \equiv \tfrac12\big(u_\alpha P_{\beta\gamma}+P_{\alpha\gamma}u_\beta\big)w_\delta , & [R]_{4,\alpha\beta\gamma\delta} & \equiv \tfrac12\big(u_\alpha Q_{\beta\delta}+Q_{\alpha\delta}u_\beta\big)v_\gamma , \notag        \\
	[R]_{5,\alpha\beta\gamma\delta} & \equiv \pi_{\alpha\beta}(\pb1) v_\gamma w_\delta ,                                  & [R]_{6,\alpha\beta\gamma\delta} & \equiv \tfrac12\big(P_{\alpha\gamma}Q_{\beta\delta}+Q_{\alpha\delta}P_{\beta\gamma}\big) , \notag \\
	[R]_{7,\alpha\beta\gamma\delta} & \equiv \pi_{\alpha\beta}(\pb1) S_{\gamma\delta} ,                                   &                                 &
	\label{eq:Rk}
\end{align}
each manifestly transverse. With these, the projected result is
\begin{align}
	\llangle TJJ\rrangle_{\rm transv.} & = \Pi^{\rm sym}_{\alpha\beta,\mu\nu}(\pb1) \pi^\rho_\gamma(\pb2) \pi^\sigma_\delta(\pb3)\times \llangle TJJ\rrangle_{\rm full} \notag        \\
	                                   & = \bar N^2\delta^{ab}\Big[\big(A_{(14)}-2A_{(16)}+A_{(26)}\big)[R]_{1,\alpha\beta\gamma\delta} \notag                                        \\
	                                   & \quad + \big(A_{(6)}-2A_{(10)}+A_{(24)}\big)[R]_{2,\alpha\beta\gamma\delta} \notag                                                           \\
	                                   & \quad + 2\big(A_{(4)}-A_{(9)}\big)[R]_{3,\alpha\beta\gamma\delta} + 2\big(A_{(12)}-A_{(15)}\big)[R]_{4,\alpha\beta\gamma\delta} \notag       \\
	                                   & \quad + A_{(13)} [R]_{5,\alpha\beta\gamma\delta} + 2A_{(2)} [R]_{6,\alpha\beta\gamma\delta} + A_{(1)} [R]_{7,\alpha\beta\gamma\delta}\Big] .
	\label{eq:intermediate13}
\end{align}
We have also checked this formula symbolically using \texttt{MATHEMATICA} by projecting each of the 28 structures in \eqref{fullcorrelator} individually.

\section{Integrals in momentum space}\label{appendB}

In section \ref{sec:calculation}, we have encountered a number of integrals over the loop momentum, the integrals $S$ and $T$. The $T$ integrals have already been solved in \cite{nastase_pre-inflationary}, and as for the $S$ integrals, $S,S_\mu,S_{\mu\nu}$ and $S_{\mu\nu\rho}$ have been solved in \cite{matheuspaper}. Below, we write the results for each of them, and in the following, we show the resolution of the scalar integral $S$ and follow the same procedure to find $S_{\mu\nu\rho\sigma}$.

According to the definitions \eqref{SandTdef}, we obtain for the integrals $S$, 
\begin{align}
S &= \frac{1}{16\pi}f_0 ,\\
S_{\mu} &= \frac{1}{16\pi}f_{\mu} ,\\
S_{\mu\nu} &= \frac{1}{16\pi}(f_{\mu\nu}+\delta_{\mu\nu}h_{0}) ,\\
S_{\mu\nu\rho} &= \frac{1}{16\pi}(\delta_{\mu\nu}h_{\rho}
  +\delta_{\mu\rho}h_{\nu}+\delta_{\nu\rho}h_{\mu}+f_{\mu\nu\rho}) ,\\
S_{\mu\nu\rho\sigma} &= \frac{1}{16\pi}\Bigl(
  -l_{0}\,\delta_{\{\mu\nu}\delta_{\rho\sigma\}}
  +\delta_{\mu\nu}h_{\rho\sigma}
  +\delta_{\mu\rho}h_{\nu\sigma}
  +\delta_{\mu\sigma}h_{\nu\rho} +\delta_{\nu\rho}h_{\mu\sigma} \nonumber \\
  & \hspace{7.3cm} +\delta_{\nu\sigma}h_{\mu\rho} +\delta_{\rho\sigma}h_{\mu\nu} +f_{\mu\nu\rho\sigma} \Bigr),
\end{align}
where, following the notation introduced in \cite{matheuspaper},
\begin{align}\label{Iabc}
    I^{(a,b,c)}&=\int_{0}^{1}dx\int_{0}^{1-x}dy\frac{x^{a}y^{b}}{\left(\Delta^{2}\right)^{c}}, \quad \text{with} \quad  \Delta^{2} \equiv p_{2}^{2}xy  +(p_{3}^{2}x+p_{1}^{2}y)(1-x-y), 
\end{align}
the remaining terms are decomposed as
\begin{equation}\label{scalars}
  f_{0}(p_1,p_2,p_3) = I^{(0,0,3/2)}, \quad h_{0}(p_1,p_2,p_3) = I^{(0,0,1/2)}, \quad l_{0}(p_1,p_2,p_3) = I^{(0,0,-1/2)},
\end{equation}
\begin{align}
  f_{\mu}(p_1,p_2,p_3) & = p_{3\mu} (I^{(0,0,3/2)}-I^{(1,0,3/2)})-p_{1\mu}I^{(0,1,3/2)},\label{fmu} \\
  \nonumber \\
  h_{\mu}(p_1,p_2,p_3) & = p_{3\mu}(I^{(0,0,1/2)}-I^{(1,0,1/2)})-p_{1\mu}I^{(0,1,1/2)}, \label{hmu} \\ 
  \nonumber \\
  f_{\mu\nu}(p_1,p_2,p_3) & = (I^{(0,0,3/2)}-2I^{(1,0,3/2)}+I^{(2,0,3/2)}) p_{3\mu}p_{3\nu} \nonumber \\ 
  & \quad +(I^{(1,1,3/2)}-I^{(0,1,3/2)}) (p_{3\mu}p_{1\nu}+p_{1\mu}p_{3\nu})+I^{(0,2,3/2)} p_{1\mu}p_{1\nu},\label{fmunu} \\
  \nonumber \\
  h_{\mu\nu}(p_1,p_2,p_3) & = p_{3\mu}p_{3\nu}(I^{(0,0,1/2)}-2I^{(1,0,1/2)}+I^{(2,0,1/2)}) \notag \\
  &\quad +(-p_{2\nu}p_{3\mu}-p_{2\mu}p_{3\nu}-2p_{3\mu}p_{3\nu})I^{(0,1,1/2)} \notag \\
  &\quad +(p_{2\mu}p_{2\nu}+p_{2\nu}p_{3\mu}+p_{2\mu}p_{3\nu}+p_{3\mu}p_{3\nu})I^{(0,2,1/2)} \notag \\
  &\quad +(p_{2\nu}p_{3\mu}+p_{2\mu}p_{3\nu}+2p_{3\mu}p_{3\nu})I^{(1,1,1/2)}, \label{hmunu} \\ 
  \nonumber \\
   f_{\mu\nu\rho}(p_1,p_2,p_3) & = (I^{(0,0,3/2)}-3I^{(1,0,3/2)}+3I^{(2,0,3/2)}-I^{(3,0,3/2)})p_{3\mu}p_{3\nu}p_{3\rho} \nonumber \\
  &\hspace{-1.2cm} -(I^{(0,1,3/2)}+I^{(2,1,3/2)}-2I^{(1,1,3/2)})(p_{3\mu}p_{3\nu}p_{1\rho}+p_{3\mu}p_{1\nu}p_{3\rho}+p_{1\mu}p_{3\nu}p_{3\rho})  \nonumber \\
  & \hspace{0.8cm} +(I^{(0,2,3/2)}-I^{(1,2,3/2)})(p_{3\mu}p_{1\nu}p_{1\rho}+p_{1\mu}p_{3\nu}p_{1\rho}+p_{1\mu}p_{1\nu}p_{3\rho}) \nonumber \\
  &\hspace{7cm}-I^{(0,3,3/2)}p_{1\mu}p_{1\nu}p_{1\rho} \, ,\\ 
  \text{and} \hspace{2.5cm}& \nonumber \\
  \hspace{-2cm} f_{\mu\nu\rho\sigma}(p_1,p_2,p_3) &= (I^{(0,0,3/2)}-4I^{(1,0,3/2)}+6I^{(2,0,3/2)}-4I^{(3,0,3/2)}+I^{(4,0,3/2)})p_{3\mu}p_{3\nu}p_{3\rho}p_{3\sigma}  \nonumber\\
  & + (-I^{(0,1,3/2)}+3I^{(1,1,3/2)}-3I^{(2,1,3/2)}+I^{(3,1,3/2)}) \nonumber\\
  & \times (p_{1\sigma}p_{3\mu}p_{3\nu}p_{3\rho}+p_{1\rho}p_{3\mu}p_{3\nu}p_{3\sigma}+p_{1\nu}p_{3\mu}p_{3\rho}p_{3\sigma}+p_{1\mu}p_{3\nu}p_{3\rho}p_{3\sigma}) \nonumber \\
  & +I^{(0,4,3/2)}p_{1\mu}p_{1\nu}p_{1\rho}p_{1\sigma}+(-I^{(0,3,3/2)}+I^{(1,3,3/2)}) \nonumber \\
  & \times (p_{1\nu}p_{1\rho}p_{1\sigma}p_{3\mu}+p_{1\mu}p_{1\rho}p_{1\sigma}p_{3\nu}+p_{1\mu}p_{1\nu}p_{1\sigma}p_{3\rho}+p_{1\mu}p_{1\nu}p_{1\rho}p_{3\sigma})  \nonumber\\
  & +(I^{(0,2,3/2)}-2I^{(1,2,3/2)}+I^{(2,2,3/2)}) \nonumber \\
  & \times \bigl(p_{1\rho}p_{1\sigma}p_{3\mu}p_{3\nu}+p_{1\nu}p_{1\sigma}p_{3\mu}p_{3\rho}+p_{1\mu}p_{1\sigma}p_{3\nu}p_{3\rho}  \nonumber\\
  & \hspace{2.7cm} +p_{1\nu}p_{1\rho}p_{3\mu}p_{3\sigma}+p_{1\mu}p_{1\rho}p_{3\nu}p_{3\sigma}+p_{1\mu}p_{1\nu}p_{3\rho}p_{3\sigma}\bigr). \label{fmunurhosigma}
\end{align}

The $T$ integrals are simpler since they reduce to the integrals of a loop with two propagators, which have been solved carefully in \cite{nastase_pre-inflationary}. They result in
\begin{align}
  T &= \frac{1}{8p_{1}}, \\
  T_{\mu} &= \frac{1}{8p_{1}}\left(\frac{1}{2}p_{1\mu}-p_{2\mu}\right), \\
  T_{\mu\nu} &= -\frac{1}{64}p_{1}\delta_{\mu\nu}+\frac{3}{64p_{1}}p_{1\mu}p_{1\nu}
  -\frac{1}{16p_{1}}p_{2\mu}p_{1\nu}-\frac{1}{16p_{1}}p_{2\nu}p_{1\mu}+\frac{1}{8p_{1}}p_{2\mu}p_{2\nu} \, .
\end{align}
These are all the integrals necessary to find the final expression for $\llangle TJJ \rrangle_{\text{flat}}$. The procedure to obtain such decompositions is described below.

\subsection{Solving $S$}

Let us take 
\begin{equation}
  S=\int\frac{d^{d}q}{\left(2\pi\right)^{d}}\times\frac{1}{q^{2}}\frac{1}{\left(q+p_{2}\right)^{2}}\frac{1}{\left(q-p_{3}\right)^{2}}.
\end{equation}
We can rewrite it using Feynman parameterization:
\begin{equation}
  \frac{1}{ABC}=2\int_{0}^{1}dx\int_{0}^{1-x}dy\frac{1}{\left[Ax+By+C\left(1-x-y\right)\right]^{3}}.
\end{equation}

If we choose $A=q^{2}$,$B=\left(q+p_{2}\right)^{2}$ and $C=\left(q-p_{3}\right)^{2}$,we have
\begin{equation}
  S=2\int_{0}^{1}dx\int_{0}^{1-x}dy\int\frac{d^{d}q}{\left(2\pi\right)^{d}}\frac{1}{\left[q^{2}x+\left(q+p_{2}\right)^{2}y+\left(q-p_{3}\right)^{2}\left(1-x-y\right)\right]^{3}}.
\end{equation}

Let us rewrite the denominator
\begin{align}
  \rightarrow & \Big[q^{2}x+\left(q+p_{2}\right)^{2}y+ \left(q-p_{3}\right)^{2}\left(1-x-y\right)\Big]^{3} \notag\\
   = & \left[q^{2}+2q\cdot\left(p_{2}y-p_{3}\left(1-x-y\right)\right)+p_{3}^{2}\left(1-x-y\right)+p_{2}^{2}y\right]^{3}.
\end{align}

Recall that, at this level, the momenta are vectorial. If we do the substitution
\begin{equation}\label{subs}
  q'_\mu=q_\mu+p_{2\mu}y-p_{3\mu}\left(1-x-y\right),
\end{equation}
the denominator becomes
\begin{align}
  \Big[q'^{2}-&\left(p_{2}y-  p_{3}\left(1-x-y\right)\right)^{2}+p_{3}^{2}\left(1-x-y\right)+p_{2}^{2}y\Big]^{3}= \notag\\
  &=\left[q'^{2}+xp_{3}^{2}(1-x)+(p_{2}+p_{3})\cdot(p_{2}+p_{3}-2p_{3}x)y-(p_{2}+p_{3})^{2}y^{2}\right]^{3},
\end{align} 
and we can use conservation of momenta $\vec{p_1} =\vec{p_2}+\vec{p_3} \Rightarrow -2p_{1}\cdot p_{3}=p_{2}^{2}-p_{1}^{2}-p_{3}^{2} $ to simplify it, so
\begin{align}
& q'^{2}+xp_{3}^{2}(1-x)+p_{1}\cdot(p_{1}-2p_{3}x)y-(p_{1})^{2}y^{2} \notag \\
& = q'^{2}+p_{2}^{2}xy+(p_{3}^{2}x+p_{1}^{2}y)(1-x-y).
\end{align}

Note that after applying momentum conservation, this integral now only depends on the magnitude of the momenta. Next, we define
\begin{equation}\label{Delta}
\Delta^{2}\equiv p_{2}^{2}xy+(p_{3}^{2}x+p_{1}^{2}y)(1-x-y),
\end{equation}
such that, finally, the scalar integral becomes
\begin{equation}
  S=2\int_{0}^{1}dx\int_{0}^{1-x}dy\int\frac{d^{d}q}{\left(2\pi\right)^{d}}\frac{1}{\left[q'^{2}+\Delta^{2}\right]^{3}}.
\end{equation}

It is useful to rewrite our integral that way because now we can use the general result in dimensional regularisation
\begin{equation}
  \int\frac{d^{d}q}{\left(2\pi\right)^{d}}\frac{1}{\left(q^{2}+\Delta^{2}\right)^{n}}=\frac{\Gamma\left(n-d/2\right)}{\left(4\pi\right)^{d/2}\Gamma\left(n\right)}\left(\Delta^{2}\right)^{d/2-n},
\end{equation}
so,
\begin{align}
  S&=2\int_{0}^{1}dx\int_{0}^{1-x}dy\frac{\Gamma\left(3-d/2\right)}{\left(4\pi\right)^{d/2}\Gamma\left(3\right)}\left(\Delta^{2}\right)^{d/2-3} \nonumber\\
  &=\frac{\Gamma\left(3-d/2\right)}{\left(4\pi\right)^{d/2}\Gamma\left(3\right)}2\int_{0}^{1}dx\int_{0}^{1-x}dy\left[p_{2}^{2}xy+(p_{3}^{2}x+p_{1}^{2}y)(1-x-y)\right]^{d/2-3}.
\end{align}

Setting $d=3$, the Gamma functions have no divergences, and we are left with an integral over the Feynman parameters,
\begin{equation}\label{Sintegral}
S=\frac{\Gamma\left(3-3/2\right)}{\left(4\pi\right)^{3/2}\Gamma\left(3\right)}2\int_{0}^{1}dx\int_{0}^{1-x}dy\left[p_{2}^{2}xy+(p_{3}^{2}x+p_{1}^{2}y)(1-x-y)\right]^{3/2-3}\equiv\frac{1}{16\pi}I^{(0,0,3/2)}.
\end{equation}
\subsection{Solving $S_{\mu\nu\rho\sigma}$}

Now we follow the same procedure for $S_{\mu\nu\rho\sigma}$. We have,
\begin{equation}
S_{\mu\nu\rho\sigma} \equiv \int\frac{d^{d}q}{\left(2\pi\right)^{d}}\frac{q_{\mu}q_{\nu}q_{\rho}q_{\sigma}}{q^{2}\left(q+p_{2}\right)^{2}\left(q-p_{3}\right)^{2}}.
\end{equation}

We use the Feynman parameterization to rewrite it as
\begin{equation}
S_{\mu\nu\rho\sigma} = 2\int_{0}^{1}dx\int_{0}^{1-x}dy\int\frac{d^{d}q}{\left(2\pi\right)^{d}}\frac{q_{\mu}q_{\nu}q_{\rho}q_{\sigma}}{\left[q^{2}x+\left(q+p_{2}\right)^{2}y+\left(q-p_{3}\right)^{2}\left(1-x-y\right)\right]^3},
\end{equation}
and apply the definitions \eqref{subs} and \eqref{Delta}, and we get
\begin{align}\label{Smunurhosigma}
S_{\mu\nu\rho\sigma} &= 2\int_{0}^{1}dx\int_{0}^{1-x}dy\int\frac{d^{d}q'}{\left(2\pi\right)^{d}}\frac{1}{\left(q'^{2}+\Delta^{2}\right)^{3}} \nonumber \\
&\quad \times \left[ \left(q'_{\mu}-yp_{2\mu}+\left(1-x-y\right)p_{3\mu}\right)\left(q'_{\nu}-yp_{2\nu}+\left(1-x-y\right)p_{3\nu}\right) \right. \nonumber \\
&\quad\quad \left. \left(q'_{\rho}-yp_{2\rho}+\left(1-x-y\right)p_{3\rho}\right)\left(q'_{\sigma}-yp_{2\sigma}+\left(1-x-y\right)p_{3\sigma}\right) \right].
\end{align}

Terms that are odd in $q_{\mu}$'s are automatically zero by dimensional regularisation, so expanding the numerator gives
\begin{align*}
&= q_{\mu}q_{\nu}q_{\rho}q_{\sigma} + \left(-yp_{2\mu}+\left(1-x-y\right)p_{3\mu}\right)\left(-yp_{2\nu}+\left(1-x-y\right)p_{3\nu}\right) \\
&\quad \qquad \times \left(-yp_{2\rho}+\left(1-x-y\right)p_{3\rho}\right)\left(-yp_{2\sigma}+\left(1-x-y\right)p_{3\sigma}\right) \\
&\quad + q_{\mu}q_{\nu}\left(-yp_{2\rho}+\left(1-x-y\right)p_{3\rho}\right)\left(-yp_{2\sigma}+\left(1-x-y\right)p_{3\sigma}\right) \\
&\quad + q_{\mu}q_{\rho}\left(-yp_{2\nu}+\left(1-x-y\right)p_{3\nu}\right)\left(-yp_{2\sigma}+\left(1-x-y\right)p_{3\sigma}\right) \\
&\quad + q_{\mu}q_{\sigma}\left(-yp_{2\nu}+\left(1-x-y\right)p_{3\nu}\right)\left(-yp_{2\rho}+\left(1-x-y\right)p_{3\rho}\right) \\
&\quad + q_{\nu}q_{\rho}\left(-yp_{2\mu}+\left(1-x-y\right)p_{3\mu}\right)\left(-yp_{2\sigma}+\left(1-x-y\right)p_{3\sigma}\right) \\
&\quad + q_{\nu}q_{\sigma}\left(-yp_{2\mu}+\left(1-x-y\right)p_{3\mu}\right)\left(-yp_{2\rho}+\left(1-x-y\right)p_{3\rho}\right) \\
&\quad + q_{\rho}q_{\sigma}\left(-yp_{2\mu}+\left(1-x-y\right)p_{3\mu}\right)\left(-yp_{2\nu}+\left(1-x-y\right)p_{3\nu}\right). 
\end{align*}
So, before we move on to the integrals on $\left(x,y\right)$, we must solve by dimensional 
regularization the three integrals over the loop momentum. The first one is the scalar integral in \eqref{Sintegral}, and the other two can be solved by tensor decomposition:
\begin{align}
&\rightarrow \int\frac{d^{d}q}{\left(2\pi\right)^{d}}\frac{q_{\mu}q_{\nu}}{\left(q^{2}+\Delta^{2}\right)^{3}}=\frac{\delta_{\mu\nu}}{2\left(4-d\right)}\frac{\Gamma\left(3-d/2\right)}{\left(4\pi\right)^{d/2}}\left(\Delta^{2}\right)^{d/2-2}\\
&\rightarrow \int\frac{d^{d}q}{\left(2\pi\right)^{d}}\frac{q_{\mu}q_{\nu}q_{\rho}q_{\sigma}}{\left(q^{2}+\Delta^{2}\right)^{3}}=\frac{1}{d\left(2+d\right)}\frac{\Gamma\left(1-d/2\right)\Gamma(2+d/2)}{\left(4\pi\right)^{d/2}\Gamma(3)\Gamma(d/2)}\left(\Delta^{2}\right)^{d/2-1}\delta_{\{\mu\nu}\delta_{\rho\sigma\}}.
\end{align}

Replacing these results into \eqref{Smunurhosigma}, we get
\begin{align}
S_{\mu\nu\rho\sigma} &= -\frac{1}{16\pi}l_{0}\left(\delta_{\mu\nu}\delta_{\rho\sigma}+\delta_{\mu\rho}\delta_{\nu\sigma}+\delta_{\mu\sigma}\delta_{\nu\rho}\right) + \frac{\delta_{\mu\nu}}{16\pi}h_{\rho\sigma} + \frac{\delta_{\mu\rho}}{16\pi}h_{\nu\sigma} + \frac{\delta_{\mu\sigma}}{16\pi}h_{\nu\rho} \nonumber \\
&\hspace{4cm}\quad + \frac{\delta_{\nu\rho}}{16\pi}h_{\mu\sigma} + \frac{\delta_{\nu\sigma}}{16\pi}h_{\mu\rho} + \frac{\delta_{\rho\sigma}}{16\pi}h_{\mu\nu} + \frac{1}{16\pi}f_{\mu\nu\rho\sigma},
\end{align}
where 
\begin{align}
&l_{0}(p_{1},p_{2},p_{3}) = \int_{0}^{1}dx\int_{0}^{1-x}dy\left(p_{2}^{2}xy+(p_{3}^{2}x+p_{1}^{2}y)(1-x-y)\right)^{1/2}, \\
&h_{\mu\nu}(p_{1},p_{2},p_{3}) = \int_{0}^{1}dx\int_{0}^{1-x}dy\frac{\left[-yp_{2\mu}+\left(1-x-y\right)p_{3\mu}\right]\left[-yp_{2\nu}+\left(1-x-y\right)p_{3\nu}\right]}{\left(p_{2}^{2}xy+(p_{3}^{2}x+p_{1}^{2}y)(1-x-y)\right)^{1/2}}, \\
&f_{\mu\nu\rho\sigma}(p_{1},p_{2},p_{3}) = \int_{0}^{1}dx\int_{0}^{1-x}dy \frac{1}{\left(p_{2}^{2}xy+(p_{3}^{2}x+p_{1}^{2}y)(1-x-y)\right)^{3/2}} \nonumber \\
&\hspace{4.5cm}\quad \times \left[-yp_{2\mu}+\left(1-x-y\right)p_{3\mu}\right]\left[-yp_{2\nu}+\left(1-x-y\right)p_{3\nu}\right] \nonumber \\
&\hspace{4.5cm}\quad \times \left[-yp_{2\rho}+\left(1-x-y\right)p_{3\rho}\right]\left[-yp_{2\sigma}+\left(1-x-y\right)p_{3\sigma}\right].
\end{align}

Then, following the notation \eqref{Iabc} of Feynmann parameterization, we rewrite them as in \eqref{scalars}\eqref{hmunu} and \eqref{fmunurhosigma}. In the next appendix, we show the analytic result for all the integrals $I^{(a,b,c)}$.

\section{Integrals over Feynman parameters}\label{appendC}

Parameterised by the notation,
\begin{align}
I^{(a,b,c)} &= \int_{0}^{1}dx\int_{0}^{1-x}dy\frac{x^{a}y^{b}}{\left(\Delta^{2}\right)^{c}} \nonumber \\
&= \int_{0}^{1}dx\int_{0}^{1-x}dy\frac{x^{a}y^{b}}{\left(p_{2}^{2}xy+(p_{3}^{2}x+p_{1}^{2}y)(1-x-y)\right)^{c}},
\end{align}
the following are all the integrals over Feynman parameters that must be solved:
\begin{center}
    \begin{table}[htbp]
        \centering
        \renewcommand{\arraystretch}{1.5}
        \begin{tabular}{@{} l c p{12.5cm} @{}}
            \toprule
            $c=3/2$ & $\rightarrow$ & $I^{(0,0,3/2)}$, $I^{(1,0,3/2)}$, $I^{(0,1,3/2)}$, $I^{(2,0,3/2)}$, $I^{(0,2,3/2)}$, 
                    $I^{(1,1,3/2)}$, $I^{(3,0,3/2)}$, $I^{(2,1,3/2)}$, $I^{(1,2,3/2)}$, $I^{(0,3,3/2)}$, 
                    $I^{(2,2,3/2)}$, $I^{(1,3,3/2)}$, $I^{(0,4,3/2)}$, $I^{(4,0,3/2)}$, $I^{(3,1,3/2)}$ \\
            
            $c=1/2$ & $\rightarrow$ & $I^{(0,0,1/2)}$, $I^{(1,0,1/2)}$, $I^{(0,1,1/2)}$, $I^{(1,1,1/2)}$, 
                    $I^{(2,0,1/2)}$, $I^{(0,2,1/2)}$ \\
            
            $c=-1/2$ & $\rightarrow$ & $I^{(0,0,-1/2)}$ \\
            \bottomrule
        \end{tabular}\label{tab:feynman_integrals}
    \end{table}
\end{center}

As noted in \cite{matheuspaper}, not all of these integrals are independent, since
\begin{equation}
I^{(a,b,c)}(p_{1},p_{2},p_{3}) = I^{(b,a,c)}(p_{3},p_{2},p_{1}),
\end{equation}
and most of them have already been solved in \cite{matheuspaper}. In the next pages, we will present the calculation of $ I^{(2,2,3/2)},I^{(3,1,3/2)}, I^{(4,0,3/2)}, I^{(2,0,1/2)}, I^{(1,1,1/2)}$ and $I^{(0,0,-1/2)}$.

Using the definition 
\begin{equation}
I^{(a,b,c)}(p_1,p_2,p_3)=\int_0^1 dx x^a I^{(b,c)}(p_1,p_2,p_3)(x),
\end{equation}
where 
\begin{align}
  I^{(b,c)}&(p_{1},p_{2},p_{3})(x)  =\int_{0}^{1-x}dy\frac{y^{b}}{\left(\alpha y^{2}+\beta(x)y+\gamma(x)\right)^{c}} ,\\
  \alpha=-p_{1}^{2},\quad\  &\beta(x)=p_{1}^{2}+x\left(p_{2}^{2}-p_{1}^{2}-p_{3}^{2}\right),\text{ and }\,\gamma(x)=(1-x)xp_{3}^{2},
\end{align}
we can first solve the integrals over $y$.

\subsection{Integrals over $y$}
The integrals over $y$ are 
\begin{itemize}
  \item $I^{(2,3/2)}$
      \begin{align}
      \hspace{-1cm} \int_{0}^{1-x} dy &\frac{y^{2}}{\big[p_{2}^{2}xy-p_{3}^{2}x(-1+x+y)-p_{1}^{2}y(-1+x+y)\big]^{3/2}} \nonumber \\
      & \hspace{2cm}=\frac{1}{p_{1}^{3}\sqrt{1-x}} \Biggl( \frac{2p_{1}(-1+x)\left[p_{1}^{2}(-1+x)-p_{2}(p_{2}+p_{3})x\right]}{p_{2}\sqrt{x}\left[-p_{1}^{2}(-1+x)+(p_{2}+p_{3})^{2}x\right]} \nonumber \\
      & \hspace{2cm}\quad + \sqrt{-1+x} \Biggl[ \coth^{-1}\left(\frac{2p_{1}p_{2}\sqrt{-1+x}\sqrt{x}}{p_{1}^{2}(-1+x)+(p_{2}-p_{3})(p_{2}+p_{3})x}\right) \nonumber \\
      &\hspace{2cm}\quad\quad - \coth^{-1}\left(\frac{2p_{1}p_{3}\sqrt{-1+x}\sqrt{x}}{p_{1}^{2}-(p_{1}^{2}-p_{2}^{2}+p_{3}^{2})x}\right) \Biggr] \Biggr) \,;
      \end{align}

  \item $I^{(0,-1/2)}$
      \begin{align}
    \hspace{-1cm}\int_{0}^{1-x}dy&\sqrt{p_{2}^{2}xy-p_{3}^{2}x(-1+x+y)-p_{1}^{2}y(-1+x+y)} \nonumber\\
    &\hspace{0.4cm}=\frac{1}{8p_{1}^{3}}\Biggl( -2p_{1}(p_{2}+p_{3})\sqrt{(1-x)x}\left[p_{1}^{2}(-1+x)+(p_{2}-p_{3})^{2}x\right] \nonumber \\
    &\hspace{0.4cm}\quad + i\left[p_{1}^{2}(-1+x)-(p_{2}-p_{3})^{2}x\right]\left[p_{1}^{2}(-1+x)-(p_{2}+p_{3})^{2}x\right] \nonumber \\
    &\hspace{0.4cm}\quad\quad \times \Bigl[ \log\left(ip_{1}(-1+x)+\frac{i(p_{2}-p_{3})(p_{2}+p_{3})x}{p_{1}}+2p_{2}\sqrt{(1-x)x}\right) \nonumber \\
    &\hspace{0.4cm}\quad\quad - \log\left(-ip_{1}(-1+x)+\frac{i(p_{2}-p_{3})(p_{2}+p_{3})x}{p_{1}}+2p_{3}\sqrt{(1-x)x}\right) \Bigr] \Biggr) \, ;
    \end{align}
  
  \item $I^{(0,1/2)}$
      \begin{align}
    &\int_{0}^{1-x} dy \frac{y^{2}}{\left[p_{2}^{2}xy-p_{3}^{2}x(-1+x+y)-p_{1}^{2}y(-1+x+y)\right]^{1/2}} \nonumber \\
    & \hspace{2.5cm}\quad = \frac{1}{2p_{1}^{3}}\Biggl( 2p_{1}(-p_{2}+p_{3})\sqrt{(1-x)x} \nonumber \\
    & \hspace{2.5cm} \quad\quad + \left[p_{1}^{2}(-1+x)+(-p_{2}^{2}+p_{3}^{2})x\right] \nonumber \\
    &\hspace{2.7cm} \quad\quad \times \Biggl[ \tan^{-1}\left(\frac{p_{1}^{2}(-1+x)+(p_{2}-p_{3})(p_{2}+p_{3})x}{2p_{1}p_{2}\sqrt{(1-x)x}}\right) \nonumber \\
    & \hspace{3cm} \quad\quad\quad\quad\quad\quad - \tan^{-1}\left(\frac{p_{1}^{2}-(p_{1}^{2}-p_{2}^{2}+p_{3}^{2})x}{2p_{1}p_{3}\sqrt{(1-x)x}}\right) \Biggr] \Biggr).
    \end{align}

\end{itemize}
\subsection{Integrals over $x$}
Now, we can calculate the full integrals by integrating over the $x$ parameter.
\begin{itemize}
  \item $I^{(2,2,3/2)}$
      \begin{align}
    I^{(2,2,3/2)}(p_{1},p_{2},p_{3}) &= \int_{0}^{1}dx \, x^{2} I^{(2,3/2)}(p_{1},p_{2},p_{3})(x) \nonumber \\
    &= \int_{0}^{1}dx \, x^{2} \frac{1}{p_{1}^{3}\sqrt{1-x}} \Biggl( \frac{2p_{1}(-1+x)\left[p_{1}^{2}(-1+x)-p_{2}(p_{2}+p_{3})x\right]}{p_{2}\sqrt{x}\left[-p_{1}^{2}(-1+x)+(p_{2}+p_{3})^{2}x\right]} \nonumber \\
    &\quad + \sqrt{-1+x} \Biggl[ \coth^{-1}\left(\frac{2p_{1}p_{2}\sqrt{-1+x}\sqrt{x}}{p_{1}^{2}(-1+x)+(p_{2}-p_{3})(p_{2}+p_{3})x}\right) \nonumber \\
    &\quad\quad - \coth^{-1}\left(\frac{2p_{1}p_{3}\sqrt{-1+x}\sqrt{x}}{p_{1}^{2}-(p_{1}^{2}-p_{2}^{2}+p_{3}^{2})x}\right) \Biggr] \Biggr).
    \end{align}

    Solving the indefinite integral and taking the lower and upper limits, we find,
    \begin{equation}
    I^{(2,2,3/2)}(p_{1},p_{2},p_{3}) = \frac{3p_{1}^{2}+4p_{1}\left(p_{2}+3p_{3}\right)+\left(p_{2}+p_{3}\right)\left(p_{2}+3p_{3}\right)}{24p_{2}\left(p_{1}+p_{2}+p_{3}\right)^{4}}\pi\, ;
    \end{equation}
  \item $I^{(3,1,3/2)}$
      \begin{align}
    I^{(3,1,3/2)}(p_{1},p_{2},p_{3}) &= \int_{0}^{1}dx \, x^{3} I^{(1,3/2)}(p_{1},p_{2},p_{3})(x) \nonumber \\
    &= \int_{0}^{1}dx \, x^{3} \frac{2}{p_{2}\sqrt{x}}\frac{\sqrt{1-x}}{p_{1}^{2}(1-x)+(p_{2}+p_{3})^{2}x}.
    \end{align}

    It amounts to
    \begin{equation}
    I^{(3,1,3/2)}(p_{1},p_{2},p_{3}) = \frac{5p_{1}^{2}+4p_{1}(p_{2}+p_{3})+(p_{2}+p_{3})^{2}}{8p_{2}\left(p_{1}+p_{2}+p_{3}\right)^{4}}\pi \, ;
    \end{equation}

  \item $I^{(4,0,3/2)}$
      \begin{align}
    \hspace{-1cm}I^{(4,0,3/2)}(p_{1},p_{2},p_{3}) &= \int_{0}^{1}dx \, x^{4} I^{(0,3/2)}(p_{1},p_{2},p_{3})(x) \nonumber \\
    &= \int_{0}^{1}dx \, x^{4} \frac{2\left(p_{2}+p_{3}\right)}{p_{2}p_{3}}\frac{1}{\sqrt{(1-x)x}\left[p_{1}^{2}(1-x)+(p_{2}+p_{3})^{2}x\right]},
    \end{align}
    that gives
    \begin{equation}
    \hspace{-1cm} I^{(4,0,3/2)}(p_{1},p_{2},p_{3}) = \frac{\left(16p_{1}^{3}+29p_{1}^{2}(p_{2}+p_{3})+20p_{1}(p_{2}+p_{3})^{2}+5(p_{2}+p_{3})^{3}\right)\pi}{8p_{2}p_{3}(p_{1}+p_{2}+p_{3})^{4}} \, ;
    \end{equation}
  
  \item $I^{(1,1,1/2)}$
  \begin{align}
      I^{(1,1,1/2)}(p_{1},p_{2},p_{3}) &= \int_{0}^{1} dx \, x I^{(1,1/2)}(p_{1},p_{2},p_{3})(x) \nonumber \\
      &= \int_{0}^{1} dx \, x \Biggl( \frac{2p_{1}(-p_{2}+p_{3})\sqrt{(1-x)x}}{2p_{1}^{3}} \nonumber \\
      &\quad + \frac{p_{1}^{2}(-1+x)+(-p_{2}^{2}+p_{3}^{2})x}{2p_{1}^{3}} \nonumber \\
      &\quad\quad \times \Biggl[ \tan^{-1}\left(\frac{p_{1}^{2}(-1+x)+(p_{2}-p_{3})(p_{2}+p_{3})x}{2p_{1}p_{2}\sqrt{(1-x)x}}\right) \nonumber \\
      &\quad\quad\quad - \tan^{-1}\left(\frac{p_{1}^{2}-(p_{1}^{2}-p_{2}^{2}+p_{3}^{2})x}{2p_{1}p_{3}\sqrt{(1-x)x}}\right) \Biggr] \Biggr).
      \end{align}

      We find,
      \begin{equation}
      I^{(1,1,1/2)}(p_{1},p_{2},p_{3}) = \frac{2p_{1}^{2}+3p_{1}(p_{2}+2p_{3})+(p_{2}+p_{3})(p_{2}+2p_{3})}{24(p_{1}+p_{2}+p_{3})^{3}}\pi \, ;
      \end{equation}

  \item $I^{(2,0,1/2)}$
      \begin{align}
    I^{(2,0,1/2)} &= \int_{0}^{1}dx \, x^{2} I^{(0,1/2)}(p_{1},p_{2},p_{3})(x) \nonumber \\
    &= \frac{i}{p_{1}}\int_{0}^{1}dx \, x^{2}\log\left(\frac{p_{1}^{2}(1-x)+2ip_{2}p_{1}\sqrt{(1-x)x}+(p_{3}^{2}-p_{2}^{2})x}{-p_{1}^{2}(1-x)+2ip_{3}p_{1}\sqrt{(1-x)x}+(p_{3}^{2}-p_{2}^{2})x}\right) \nonumber \\
    &= \frac{\left(8p_{1}^{2}+9p_{1}\left(p_{2}+p_{3}\right)+3\left(p_{2}+p_{3}\right)^{2}\right)\pi}{24\left(p_{1}+p_{2}+p_{3}\right)^{3}};
    \end{align}

  \item $I^{(0,0,-1/2)}$
      \begin{align}
    I^{(0,0,-1/2)} &= \int_{0}^{1}dx \, I^{(0,-1/2)}(p_{1},p_{2},p_{3})(x) \nonumber \\
    &= \int_{0}^{1}dx \, \frac{1}{8p_{1}^{3}}\Biggl( -2p_{1}(p_{2}+p_{3})\sqrt{(1-x)x}\left[p_{1}^{2}(-1+x)+(p_{2}-p_{3})^{2}x\right] \nonumber \\
    &\quad + i\left[p_{1}^{2}(-1+x)-(p_{2}-p_{3})^{2}x\right]\left[p_{1}^{2}(-1+x)-(p_{2}+p_{3})^{2}x\right] \nonumber \\
    &\quad\quad \times \Bigl[ \log\left(ip_{1}(-1+x)+\frac{i(p_{2}-p_{3})(p_{2}+p_{3})x}{p_{1}}+2p_{2}\sqrt{(1-x)x}\right) \nonumber \\
    &\quad\quad\quad - \log\left(-ip_{1}(-1+x)+\frac{i(p_{2}-p_{3})(p_{2}+p_{3})x}{p_{1}}+2p_{3}\sqrt{(1-x)x}\right) \Biggr] \Biggr). \nonumber 
    \end{align}
    This results in
    \begin{equation}
    \frac{\left(p_{1}^{3}+2p_{1}^{2}\left(p_{2}+p_{3}\right)+2p_{1}\left(p_{2}^{2}+p_{2}p_{3}+p_{3}^{2}\right)+\left(p_{2}+p_{3}\right)\left(p_{2}^{2}+p_{2}p_{3}+p_{3}^{2}\right)\right)}{24\left(p_{1}+p_{2}+p_{3}\right)^{2}}\pi.
    \end{equation}
\end{itemize}
\subsection{List of results}\label{append_list}
In summary, the analytical results for all scalar integrals appearing in the decomposition of $\llangle TJJ \rrangle_{\text{flat}}$ are
\begin{align}
  I^{(0,0,3/2)} &= \frac{2\pi}{p_{1}p_{2}p_{3}}, \hspace{5cm} I^{(0,0,1/2)} = \frac{\pi}{p_{1}+p_{2}+p_{3}},\notag \\
  I^{(1,0,3/2)} &= \frac{2\pi}{p_{2}p_{3}(p_{1}+p_{2}+p_{3})}, \hspace{3.2cm} I^{(1,1,3/2)} = \frac{\pi}{p_{2}(p_{1}+p_{2}+p_{3})^{2}}, \notag \\
  I^{(2,0,3/2)} &= \frac{2p_{1}+p_{2}+p_{3}}{p_{2}p_{3}(p_{1}+p_{2}+p_{3})^{2}}\pi,  \hspace{2.8cm} I^{(1,0,1/2)} = \frac{(2p_{1}+p_{2}+p_{3})}{4(p_{1}+p_{2}+p_{3})^{2}}\pi \notag \\ 
I^{(2,1,3/2)} & = \frac{3p_{1}+p_{2}+p_{3}}{4p_{2}(p_{1}+p_{2}+p_{3})^{3}}\pi, \notag\\
  I^{(3,0,3/2)} & = \frac{8p_{1}^{2}+9(p_{2}+p_{3})p_{1}+3(p_{2}+p_{3})^{2}}{4p_{2}p_{3}(p_{1}+p_{2}+p_{3})^{3}}\pi,  \notag \\
  I^{(2,0,1/2)} & = \frac{\left(8p_{1}^{2}+9p_{1}(p_{2}+p_{3})+3(p_{2}+p_{3})^{2}\right)}{24(p_{1}+p_{2}+p_{3})^{3}}\pi, \notag \\
  I^{(3,1,3/2)} & = \frac{5p_{1}^{2}+4p_{1}(p_{2}+p_{3})+(p_{2}+p_{3})^{2}}{8p_{2}(p_{1}+p_{2}+p_{3})^{4}}\pi, \notag\\
  I^{(1,1,1/2)} &= \frac{2p_{1}^{2}+3p_{1}(p_{2}+2p_{3})+(p_{2}+p_{3})(p_{2}+2p_{3})}{24(p_{1}+p_{2}+p_{3})^{3}}\pi, \notag\\
  I^{(2,2,3/2)} &= \frac{3p_{1}^{2}+4p_{1}(p_{2}+3p_{3})+(p_{2}+p_{3})(p_{2}+3p_{3})}{24p_{2}(p_{1}+p_{2}+p_{3})^{4}}\pi,\notag \\
  I^{(4,0,3/2)} &= \frac{16p_{1}^{3}+29p_{1}^{2}(p_{2}+p_{3})+20p_{1}(p_{2}+p_{3})^{2}+5(p_{2}+p_{3})^{3}}{8p_{2}p_{3}(p_{1}+p_{2}+p_{3})^{4}}\pi, \notag\\
  I^{(0,0,-1/2)} &= \frac{p_{1}^{3}+2p_{1}^{2}(p_{2}+p_{3})+2p_{1}(p_{2}^{2}+p_{2}p_{3}+p_{3}^{2})+(p_{2}+p_{3})(p_{2}^{2}+p_{2}p_{3}+p_{3}^{2})}{24(p_{1}+p_{2}+p_{3})^{2}}\pi.\label{intfeynman}
\end{align}

\section{Regularization and Inversion of Transverse Two-Point Functions}\label{append_inversion}

In this appendix, we give a "physicist's derivation" of the \emph{Moore–Penrose pseudo-inverse} we use to define the 
\begin{equation}
    \mathcal{L}_{\text{Maxwell}} = \frac{1}{4}F_{\mu\nu}F^{\mu\nu} + \frac{1}{2\xi}(\partial_\mu A^\mu)^2.
\end{equation}

In momentum space, the kinetic operator is
\begin{equation}
    \mathcal{O}^{\mu\nu}(k) = k^2 g^{\mu\nu} - \left(1 - \frac{1}{\xi}\right)k^\mu k^\nu.
\end{equation}

We decompose the metric into its transverse and longitudinal projectors. Let $\pi^{\mu\nu} = g^{\mu\nu} - 
\frac{k^\mu k^\nu}{k^2}$ be the transverse projector. Substituting $g^{\mu\nu} = \pi^{\mu\nu} + \frac{k^\mu 
k^\nu}{k^2}$ into the kinetic operator yields:
\begin{align}
    \mathcal{O}^{\mu\nu}(k) &= k^2 \left(\pi^{\mu\nu} + \frac{k^\mu k^\nu}{k^2}\right) - k^\mu k^\nu + 
    \frac{k^2}{\xi}\frac{k^\mu k^\nu}{k^2} \notag \\
    &= k^2 \pi^{\mu\nu} + \frac{k^2}{\xi}\frac{k^\mu k^\nu}{k^2}.
\end{align}
In this basis, the matrix is diagonalized into its transverse and longitudinal subspaces. The transverse 
eigenvalue is $k^2$, and the longitudinal eigenvalue is $k^2/\xi$. 

The photon propagator $D_{\nu\rho}(k)$ is defined as the right-inverse of the kinetic operator: $
\mathcal{O}^{\mu\nu} D_{\nu\rho} = \delta^\mu_\rho$. Because the projectors are orthogonal ($
\pi^{\mu\nu} k_\nu = 0$) and summing to the identity, the inverse matrix is simply constructed by inverting the 
respective eigenvalues:
\begin{equation}
    D_{\mu\nu}(k) = \frac{1}{k^2} \pi_{\mu\nu} + \frac{\xi}{k^2} \frac{k_\mu k_\nu}{k^2}.
\end{equation}

We now take the Landau gauge limit, $\xi \to 0$. Physically, the gauge-fixing term in the action $\frac{1}
{2\xi}(\partial_\mu A^\mu)^2$ blows up, meaning we need an infinite amount of energy to excite 
longitudinal fluctuations ($\partial_\mu A^\mu \neq 0$). This translates into the fact that the longitudinal 
eigenvalue of the operator $\mathcal{O}^{\mu\nu}$ diverges to infinity, and the corresponding eigenvalue 
of the \textit{inverse} matrix $D_{\mu\nu}$ vanishes:
\begin{equation}
    \lim_{\xi \to 0} D_{\mu\nu}(k) = \frac{1}{k^2} \pi_{\mu\nu}.
\end{equation}
By imposing $\xi =0$ at operator level, the propagator is confined to the transverse subspace.

We can do the same to obtain the formal inversion of the two-point function of the global conserved 
current $J_\mu(p)$. Since current is conserved ($p^\mu J_\mu = 0$), the two-point function is strictly 
transverse:
\begin{equation}
    G_{\mu\nu}(p) \equiv \langle J_\mu(p) J_\nu(-p) \rangle = A_J(p) \pi_{\mu\nu}(p)\;,
\end{equation}
where $A_J(p)$ is the scalar form factor. 

Unlike the gauge field, $J_\mu$ has no gauge redundancy; its transversality is a physical constraint (a Ward identity). Consequently, $G_{\mu\nu}$ is a strictly singular matrix with a longitudinal eigenvalue of 
exactly zero. We define its inverse by introducing a regulator $\lambda$ to shift the zero eigenvalue:
\begin{equation}
    M_{\mu\nu}(p) \equiv  A_J(p) \pi_{\mu\nu}(p) + \lambda \frac{p_\mu p_\nu}{p^2}.
\end{equation}
The matrix $M_{\mu\nu}$ is non-singular for any $\lambda \neq 0$. We demand an inverse $
(M^{-1})^{\nu\rho}$ such that $M_{\mu\nu}(M^{-1})^{\nu\rho} = \delta_\mu^\rho$. Utilizing the orthogonality 
of the projectors, we invert the coefficients identically to the QED derivation:
\begin{equation}
    (M^{-1})^{\mu\nu}(p) = \frac{1}{A_J(p)} \pi^{\mu\nu}(p) + \frac{1}{\lambda} \frac{p^\mu p^\nu}{p^2}.
\end{equation}
To recover the physically relevant inverse, we must remove the longitudinal modes introduced by the mathematical regulator. Just as we 
did for the photon propagator, we force the longitudinal component of the inverse to vanish by taking the 
limit $\lambda \to \infty$. This implies that these longitudinal modes we artificially introduce are suppressed from the physical inverse of the correlator:
\begin{equation}
    \tilde{G}^{\mu\nu}(p) \equiv \lim_{\lambda \to \infty} (M^{-1})^{\mu\nu}(p) = \frac{1}{A_J(p)} \pi^{\mu\nu}
    (p).
\end{equation}
The regularized limit derived above explicitly constructs the Moore-Penrose pseudoinverse of the original 
singular tensor. 
For a generic singular matrix $G$, the Moore-Penrose 
pseudoinverse $G^+$ is defined by $G G^+ G = G$. We can test that the inverse correlator we derived 
satisfies
\begin{equation}
    G_{\mu\alpha} \tilde{G}^{\alpha\beta} G_{\beta\nu} = \left[ A_J \pi_{\mu\alpha} \right] \left[ \frac{1}{A_J} 
    \pi^{\alpha\beta} \right] \left[ A_J \pi_{\beta\nu} \right].
\end{equation}
Since the scalar factors cancel trivially and projection operators are idempotent ($\pi_{\mu\alpha} 
\pi^{\alpha\beta} = \pi_\mu^\beta$), the contraction reduces to:
\begin{equation}
    A_J \left( \pi_\mu^\beta \pi_{\beta\nu} \right) = A_J \pi_{\mu\nu} = G_{\mu\nu}.
\end{equation}

This ensures that contractions with external bulk fields are evaluated correctly over the non-singular 
physical transverse subspace.

\section{Form factors}\label{append_formfactors}

Here we list the expressions found for the form factors on the decompositions of $\llangle TJJ \rrangle_{\text{full}}$  in \eqref{fullcorrelator}. These results were obtained by replacing the previous solutions for the scalar integrals and using \texttt{MATHEMATICA} to organize the tensor structure.

\begin{align*}
A_{(1)} &= \frac{\bar{p}_1^2 \bar{p}_2 + \bar{p}_1 \bar{p}_2^2 - 5 \bar{p}_1^2 P + \bar{p}_2^2 P + \bar{p}_3 P^2}{6 P^2} ,\\
A_{(2)} &= \frac{(2 \bar{p}_1 + \bar{p}_2 + \bar{p}_3) (\bar{p}_1^2 - \bar{p}_2^2 - 4 \bar{p}_2 \bar{p}_3 - \bar{p}_3^2)}{12 P^2} ,\\
A_{(3)} &= \frac{3 (\bar{p}_1 - \bar{p}_2) (\bar{p}_1 + \bar{p}_2)^2 + 3 (\bar{p}_1 + \bar{p}_2) (3 \bar{p}_1 + \bar{p}_2) \bar{p}_3 + (9 \bar{p}_1 + 5 \bar{p}_2) \bar{p}_3^2 + 3 \bar{p}_3^3}{12 \bar{p}_2 P^3} ,\\
A_{(4)} &= -\frac{\bar{p}_3^2}{3 P^3}, \quad A_{(5)} = -\frac{\bar{p}_3 (3 \bar{p}_1 (\bar{p}_1 + \bar{p}_2) + (3 \bar{p}_1 + \bar{p}_2) \bar{p}_3)}{3 \bar{p}_2 P^3} ,\\
A_{(6)} &= \frac{3 (\bar{p}_1 + \bar{p}_2)^2 + 9 (\bar{p}_1 + \bar{p}_2) \bar{p}_3 + 4 \bar{p}_3^2}{6 P^3}, \quad A_{(7)} = \frac{(\bar{p}_1 + \bar{p}_2) \bar{p}_3^2}{\bar{p}_1 \bar{p}_2 P^4} ,\\
A_{(8)} &= \frac{-\bar{p}_1^2 + \bar{p}_2^2 + \bar{p}_3^2}{6 P^3}, \quad A_{(9)} = -\frac{2 \bar{p}_1^2 + \bar{p}_2^2 + 3 \bar{p}_2 \bar{p}_3 + 4 \bar{p}_3^2 + 3 \bar{p}_1 (\bar{p}_2 + 2 \bar{p}_3)}{6 P^3} ,\\
A_{(10)} &= \frac{7 \bar{p}_1^2 + 5 \bar{p}_2^2 + 12 \bar{p}_2 \bar{p}_3 + 5 \bar{p}_3^2 + 12 \bar{p}_1 (\bar{p}_2 + \bar{p}_3)}{6 P^3} ,\\
A_{(11)} &= \frac{-\bar{p}_1 (\bar{p}_1 + \bar{p}_2)^2 + (\bar{p}_1 + 2 \bar{p}_2) \bar{p}_3^2}{2 \bar{p}_1 \bar{p}_2 P^4}, \quad A_{(12)} = -\frac{2 (\bar{p}_1 + \bar{p}_2) (\bar{p}_1 + 2 \bar{p}_2) + 3 (\bar{p}_1 + \bar{p}_2) \bar{p}_3 + \bar{p}_3^2}{6 P^3} ,\\
A_{(13)} &= \frac{7 \bar{p}_1^2 + 9 \bar{p}_1 (\bar{p}_2 + \bar{p}_3) + 2 (\bar{p}_2^2 + 3 \bar{p}_2 \bar{p}_3 + \bar{p}_3^2)}{6 P^3} ,\\
A_{(14)} &= -\frac{(\bar{p}_1 + \bar{p}_2)^2 + 4 (\bar{p}_1 + \bar{p}_2) \bar{p}_3 + \bar{p}_3^2}{2 \bar{p}_1 P^4}, \quad A_{(15)} = -\frac{\bar{p}_2^2}{3 P^3},\\
A_{(16)} &= -\frac{7 \bar{p}_1^2 + 10 \bar{p}_1 (\bar{p}_2 + \bar{p}_3) + 3 (\bar{p}_2^2 + 4 \bar{p}_2 \bar{p}_3 + \bar{p}_3^2)}{6 \bar{p}_1 P^4} ,\\
A_{(17)} &= -\frac{(3 \bar{p}_1 (\bar{p}_1 + \bar{p}_2) + (3 \bar{p}_1 + \bar{p}_2) \bar{p}_3) (-\bar{p}_1^2 + \bar{p}_2^2 + \bar{p}_3^2)}{6 \bar{p}_2 \bar{p}_3 P^3} ,\\
A_{(18)} &= -\frac{(\bar{p}_1 + \bar{p}_2) (\bar{p}_1^2 - \bar{p}_2^2 - \bar{p}_3^2)}{2 \bar{p}_1 \bar{p}_2 P^4}, \quad A_{(19)} = \frac{1}{4 \bar{p}_3} + \frac{-2 \bar{p}_2^2 + 3 \bar{p}_2 P - 3 P^2}{6 P^3}, \\
A_{(20)} &= -\frac{6 \bar{p}_1^4 + 12 \bar{p}_1^3 (\bar{p}_2 + \bar{p}_3) + 4 \bar{p}_1 \bar{p}_2 \bar{p}_3 (\bar{p}_2 + \bar{p}_3) - 3 \bar{p}_2 \bar{p}_3 (\bar{p}_2^2 + \bar{p}_3^2) + \bar{p}_1^2 (6 \bar{p}_2^2 + 19 \bar{p}_2 \bar{p}_3 + 6 \bar{p}_3^2)}{6 \bar{p}_1 \bar{p}_2 \bar{p}_3 P^4} ,\\
A_{(21)} &= -\frac{\bar{p}_2 (3 \bar{p}_1 (\bar{p}_1 + \bar{p}_2) + (3 \bar{p}_1 + \bar{p}_2) \bar{p}_3)}{3 \bar{p}_3 P^3}, \quad A_{(22)} = \frac{\bar{p}_2 (\bar{p}_1 + \bar{p}_2)}{\bar{p}_1 P^4} - \frac{2}{3 P^3}, \\
A_{(23)} &= -\frac{\bar{p}_1^3 + 2 \bar{p}_1^2 \bar{p}_3 - 2 \bar{p}_2^2 \bar{p}_3 + \bar{p}_1 (-\bar{p}_2^2 + \bar{p}_3^2)}{2 \bar{p}_1 \bar{p}_3 P^4} ,\\
A_{(24)} &= \frac{3 \bar{p}_1^2 + 9 \bar{p}_1 \bar{p}_2 + 4 \bar{p}_2^2 + 6 \bar{p}_1 \bar{p}_3 + 9 \bar{p}_2 \bar{p}_3 + 3 \bar{p}_3^2}{6 P^3} ,\\
A_{(25)} &= \frac{-2 \bar{p}_1 (\bar{p}_1 + \bar{p}_2) + \bar{p}_1 \bar{p}_3 + 3 \bar{p}_3^2}{3 \bar{p}_1 P^4}, \quad A_{(26)} = -\frac{\bar{p}_1^2 + \bar{p}_2^2 + 4 \bar{p}_2 \bar{p}_3 + \bar{p}_3^2 + 2 \bar{p}_1 (2 \bar{p}_2 + \bar{p}_3)}{2 \bar{p}_1 P^4} ,\\
A_{(27)} &= -\frac{(\bar{p}_1 + \bar{p}_3)(\bar{p}_1^2 - \bar{p}_2^2 - \bar{p}_3^2)}{2 \bar{p}_1 \bar{p}_3 P^4}, \quad A_{(28)} = \frac{\bar{p}_2^2 (\bar{p}_1 + \bar{p}_3)}{\bar{p}_1 \bar{p}_3 P^4}
\end{align*}

\bibliography{HoloCosmoTJJ}
\bibliographystyle{utphys}
\end{document}